\documentclass[twocolumn]{aastex62}
\usepackage{natbib}
\usepackage{apjfonts} 
\usepackage{xcolor}
\usepackage{multirow}
\newcommand{\msun}{$\rm M_{\sun}$}
\newcommand{\rsun}{$\rm R_{\sun}$}
\newcommand{\Lsun}{$\rm L_{\sun}$}
\newcommand{\ysun}{$\rm Y_{\sun}$}
\newcommand{\zsun}{$\rm Z_{\sun}$}
\newcommand{\Mzams}{$\rm M_{ZAMS}$}
\newcommand{\Teff}{$\rm T_{eff}$}

\newcommand{\e}[2]{$\rm #1 \times 10^{#2}$}
\newcommand{\powten}[1]{$\rm 10^{#1}$}
\newcommand{\iso}[2]{$\rm ^{#1}#2$}

\accepted{2019 November 26}
\submitjournal{ApJ}

\shorttitle{Type IIP SN progenitors \& dust obscuration}
\shortauthors{Wagle \& Ray}

\begin{document}

\title{Type IIP Supernova Progenitors II: Stellar Mass and Obscuration by the Dust in the Circumstellar Medium}

\correspondingauthor{Gururaj A. Wagle}
\email{guru.w84@gmail.com}

\author[0000-0002-3356-5855]{Gururaj A. Wagle}
\affiliation{Homi Bhabha Centre for Science Education - Tata Institute of Fundamental Research, Mankhurd, Mumbai 400088, India}

\author[0000-0003-2404-0018]{Alak Ray}
\affiliation{Homi Bhabha Centre for Science Education - Tata Institute of Fundamental Research, Mankhurd, Mumbai 400088, India}


\begin{abstract}

It has been well established from a variety of observations that red supergiants (RSGs) loose a lot of mass in stellar wind. \deleted{Since material in the wind can condense into dust, the }\added{Dust formed in this} emitted gas over a few decades before core-collapse can lead to substantial extinction and obscure the intrinsic luminosity of the progenitor RSG. This may lead to a difficulty in determining the range of progenitor masses that lead to the different classes of supernovae. Even the \deleted{close}\added{nearby}, well studied supernovae with pre-explosion observations, such as SN 2013ej may \deleted{still} suffer from this uncertainty in the progenitor mass. We explore \added{here two different}\deleted{ the range of} masses proposed for its progenitor\added{. We compute their pre-supernova characteristics} using Modules for Experiments in Stellar Astrophysics (MESA). We show that a non-rotating star with the initial mass of 26 M$_{\odot}$ \deleted{ may end up with a substantially smaller pre-supernova mass even with moderate mass loss rate. It} would require a considerable amount of circumstellar medium \added{(A$_V \sim$ 3)} to obscure its high luminosity \deleted{to be constrained within}\added{given} the observed pre-explosion magnitudes detected by the Hubble Space Telescope (HST). Such a high value of visual extinction appears to be inconsistent with that derived for SN 2013ej as well as SN 2003gd in the same host galaxy M74. In contrast, the evolutionary models of lower mass (13 M$_{\odot}$) star are easily accommodated within the observed HST magnitudes. Some of the 26 M$_{\odot}$ simulations show luminosity variation in the last few years which could be discriminated by high cadence and multiband monitoring of supernova candidates in nearby galaxies.\deleted{ Since presupernova modeling is crucial for a simulation of the explosion of a supernova, we explore a variety of model parameters and compare outcomes of variables that may determine the possibility of explosion of the progenitor.} We demonstrate that our calculations are well resolved with adequate zoning and evolutionary time-steps.
\end{abstract}



\section{Introduction} \label{sec:intro}

Core-collapse Supenovae (CCSNe) occur as a result of gravitational collapse of massive stars (typically, $>$ 8 \msun \ at Zero Age Main Sequence, ZAMS) at the end of their evolution. \added{CCSNe are classified based on the observations of optical and infrared light-curves and spectra. The eponymous plateau in SN IIP visible band light curves typically lasts for 60-100 rest-frame days after which exponentially decaying tails follow\footnote{In contrast to IIP, the light-curves of IIL (linear) are similar to SNe Type I. While SNe Type II show hydrogen lines in their spectra, SNe Type I show no obvious hydrogen lines. In addition to the well-known photometric subclasses IIP and IIL, there exists a spectroscopic subclass IIn which is distinguished by relatively narrow emission lines and slowly declining light-curves.} \citep[for details on classification of SNe, see the review article by][]{Filippenko:1997aa}.} A volume limited sample of CCSNe within 60 Mpc \citep{Smith:2011aa} shows that nearly 48\% of these are supernovae of Type IIP, which have progenitors with a large hydrogen rich envelope at the time of their explosion. Because of the observational constraints, the progenitors corresponding to supernovae can be identified only for relativity nearby cases (at $d \leq 30 \rm \; Mpc$). \object{SN 2013ej}, a type IIP SN like its predecessor \object{SN 2003gd}, occurred in the same host galaxy M74 at a distance of 9.0$^{+0.4}_{-0.6}$ Mpc \citep{Dhungana:2016aa} and was followed extensively by many groups in UV, optical, infrared bands \citep{Yuan:2016aa,Huang:2015aa,Bose:2015aa,Richmond:2014aa,Fraser:2014aa,Valenti:2014aa} and in X-ray bands \citep{Chakraborti:2016aa}. The estimated mass of the progenitor of SN 2013ej in the literature ranges between 11--16 \msun, while the X-ray measurement of \citet{Chakraborti:2016aa} points to a ZAMS mass of 13.7 \msun \ for the progenitor. \citet{Das:2017aa} restrict the ZAMS mass of the progenitor star of SN 2013ej using the archival HST observations and their simulations to an upper bound of 14 \msun . They also show through their 1-D simulations of evolution of a 13 \msun \ star and its explosion that the observed light curves are fitted well with their simulations, when they included enhanced mass loss towards the end stages of evolution of the star. On the other hand, \citet{Utrobin:2017aa} argue for a significantly higher progenitor mass of 25.5--29.5 \msun \ on the main sequence (and the ejecta mass of 23--26 \msun) based on arguments of high velocity $\rm Ni^{56}$ ejecta. They also claim that a sufficient mass loss rate can produce a circum-stellar envelope at a distance about \powten{15} cm that would hide the pre-SN light from such a massive progenitor star.

We present here a few cases of an isolated (single), non-rotating and non-magnetized star as a progenitor of SN 2013ej for two fixed ZAMS masses of 13 \msun \ and 26 \msun . In the companion Paper I \citep{Wagle:2019aa}, we have studied in detail the effect of convective overshoot on the evolution and explodibility of the progenitor for ZAMS mass of 13 \msun . \deleted{In this work, we show the effects of variation of model parameters such as mass resolution, isotopes network size, mass loss in stellar wind, etc. on the structure of progenitor through the history of the star's evolution and the core-collapse (CC) stage. For this study, we use a 1-D stellar evolution code Modules for Experiments in Stellar Astrophysics \citep[MESA,][]{Paxton:2011aa,Paxton:2013aa,Paxton:2015aa,Paxton:2018aa} to evolve the star from pre-MS through CC. }\added{Here, we also}\deleted{We} test the case of a 26 \msun \ star as a possible progenitor of SN 2013ej. We study its luminosity variations in the late stages of stellar evolution and calculate its visual extinction due to dust formation in the  circumstellar medium (CSM) formed by mass loss through stellar wind and compare these with archival observations of the progenitor star. \deleted{We also show that the mass resolution has considerable effect on the evolution of 26 \msun \ models in the Hertzsprung-Russell diagram (HRD). }We also demonstrate that our models have adequately fine mass resolution\deleted{ with the above choices of \textit{dm}. In our future work, we will simulate the explosion of these models through 1-D SN explosion codes and compare with the observed light curves and expansion velocity profiles to test the viability of the models. We also present here a slightly modified version of the ``test suite"  for a pre-SN star that accompanies the standard MESA distribution for both the above ZAMS masses and compare its output with the models with better mass and temporal resolution controls, as well as larger network sizes.}  In our future work, we will simulate the explosion of these models through 1-D SN explosion codes and compare with the observed light curves and expansion velocity profiles to test the viability of the models.

In section \ref{sec:methods}, we describe the methods of computational simulations and the stellar evolution using MESA. In section \ref{sec:results}, we discuss the results of variation of different MESA parameters on the stellar structure for our models. We also show that our models have adequate mass resolution. We discuss the extinction due to dust (formed in the CSM) on the observed magnitudes of the progenitor star. In section \ref{sec:conclusion}, we discuss and summarize our conclusions.

\begin{splitdeluxetable*}{cccccccclBccccchh}
\tablecaption{Inlist parameters \& MESA predicted core properties for  \Mzams \ = 13 \msun \ \& 26 \msun \ models with Z = 0.006 and $f$ = 0.025
\label{tab:13m_26m_param}}
\tablewidth{0pt}
\tablehead{
\colhead{\textit{dm}} & \colhead{overshoot} &
\multicolumn{3}{c}{delta$\_$lg$\_$X$\_$cntr$\_$limit/hardlimit} &
\colhead{Max \textit{dt}} & \colhead{Dutch Wind} & \colhead{Network} & 
\colhead{model id}  &
\colhead{$\rm He_{core}$} & \colhead{$\rm C_{core}$} & 
\colhead{$\rm O_{core}$} & \colhead{$\rm Si_{core}$} & 
\colhead{$\rm Fe_{core}$} & \nocolhead{$\xi_{2.5}$} & 
\nocolhead{$\xi_{1.6}$} \\[-1.5ex]
\colhead{($\rm M_{\odot}$)} & \colhead{parameter, $f_0$} & 
\colhead{Ne} & \colhead{O} & \colhead{Si} &
\colhead{Change} & \colhead{$\eta$} & \colhead{size} &
\colhead{} & \colhead{} &
\colhead{} & \colhead{} &
\colhead{} & \nocolhead{} &
\nocolhead{}
}
\colnumbers
\startdata
\multicolumn{9}{c}{M$_{ZAMS}$ = 13 $\rm M_{\odot}$} & \multicolumn{7}{c}{M$_{ZAMS}$ = 13 $\rm M_{\odot}$} \\
\hline
0.007 & 0.050 & 0.02/0.03 & 0.02/0.03 & 0.02/0.03 & 1.15 & 0.5 & 22 isotopes &  13$\rm M_{\odot}$ \ model 1\phn & 
 4.271 & 2.108 & 2.056 & 1.576 & 1.477 & 0.029 & 0.770 \\
0.007 & 0.005 & 0.02/0.03 & 0.02/0.03 & 0.02/0.03 & 1.15 & 0.5 & 22 isotopes & 13$\rm M_{\odot}$ \ model 2\phn &
4.415 & 2.179 & 2.120 & 1.611 & 1.510 & 0.035 & 0.939 \\
0.007 &  0.005 & 0.02/0.03 & 0.02/0.03 & 0.02/0.03 & 1.15 & 0.5 & 79 isotopes\tablenotemark{a} & 13$\rm M_{\odot}$ \ model 3\phn & 
 4.402 & 2.187 & 1.898 & 1.545 & 1.481 & 0.030 & 0.561 \\
0.01 & 0.050 & 0.02/0.03 & 0.02/0.03 & 0.02/0.03 & 1.15 & 1.0 & 22 isotopes & 13$\rm M_{\odot}$ \  model 4\phn &
4.260 & 2.102 & 2.052 & 1.591 & 1.495 & -- & -- \\
0.007 &  0.050 & 0.02/0.03 & 0.02/0.03 & 0.02/0.03 & 1.2/1.15 & 0.5 & 45/204 isotopes & 13$\rm M_{\odot}$ \ model 5\tablenotemark{b} &
4.265 & 2.059 & 1.929 & 1.497 & 1.365 &  -- & -- \\
-- &   0.005 & -- & -- & -- & 1.2\phn & 0.5 & 21 isotopes & 13$\rm M_{\odot}$ \  modified test suite & 
4.381 & 2.147 & 1.969 & 1.593 & 1.468 & 0.034 & 0.721  \\
\hline
\multicolumn{9}{c}{M$_{ZAMS}$ = 26 $\rm M_{\odot}$} & \multicolumn{7}{c}{M$_{ZAMS}$ = 26 $\rm M_{\odot}$} \\
\hline
0.007 & 0.050 & 0.02/0.03 & 0.02/0.03 & 0.02/0.03 & 1.2\phn & 0.5 & 22 isotopes & 26$\rm M_{\odot}$ \ model 1\phn &
8.406 & 4.610 & 1.889 & 0.000 & 1.529 & 0.128 & 0.967 \\
0.01 & 0.050 & 0.015/0.03 & 0.015/0.03 & 0.015/0.03 & 1.2\phn & 0.5 & 22 isotopes & 26$\rm M_{\odot}$ \ model 2\tablenotemark{c} &
6.876 & 3.262 & 2.948 & 1.729 & 1.589 & 0.187 & 1.086 \\
0.01 & 0.020 & 0.01/0.02 & 0.01/0.02 & 0.01/0.02 & 1.2\phn & 0.5 & 22 isotopes & 26$\rm M_{\odot}$ \ model 3\phn &
7.572 & 3.952 & 3.820 & 1.690 & 1.538 & 0.151 & 1.067 \\
0.01 & 0.050 & 0.01/0.02 & 0.02/0.04 & 0.01/0.02 & 1.2\phn & 0.5 & 79 isotopes & 26$\rm M_{\odot}$ \ model 4\tablenotemark{c} &
6.876 & 3.253 & 2.963 & 1.811 & 1.587 & -- & -- \\
-- & 0.050 & -- & -- & -- & 1.2\phn & 0.5 & 21 isotopes & 26$\rm M_{\odot}$ \ modified test suite &
9.209 & 5.294 & 2.427 & 1.660 & 1.508 & 0.138 & 0.874 \\
\enddata
\tablenotetext{a}{Network provided by \citet{Farmer:2016aa} at MESA market place.}
\tablenotetext{b}{max$\_$timestep$\_$factor was changed after TAMS from 1.2 to 1.15. The network was changed from 45 (mesa$\_$45.net) to 203 isotopes network \citep[si$\_$burn.net, provided by][]{Renzo:2017aa}, and the maximum numbers of grid points allowed was changed to 5000 from MESA default value of 8000, after O-depletion \citep[ref.][for the definition of O-depletion and the explanation for these changes.]{Renzo:2017aa}}
\tablenotetext{c}{26 $\rm M_{\odot}$ \ model 2 run was restarted at model number 7000 by changing delta$\_$lg$\_$X$\_$cntr$\_$limit values for Ne,O \& Si to 0.015 (and corresponding hard limit to 0.03) and model 4 run was restarted at model number 6000 by changing only the delta$\_$lg$\_$XNe$\_$cntr$\_$limit to 0.02 (hard limit to 0.04) from the original values of corresponding limit of 0.01 (hardlimit of 0.02). The model was running into convergence problems with the original lg$\_$XNe$\_$cntr$\_$limit values at those stages. In other models, the values were kept constant through CC}
\tablecomments{\texttt{delta$\_$lg$\_$X$\_$cntr$\_$limit} \& \texttt{hardlimit} values were set to 0.01 \& 0.02, respectively through CC for H, He and C in all of the above models. See section \ref{sec:methods} \deleted{and appendix \ref{app:inlists} }for more details on the choices of other parameters\deleted{ for the models} listed here.}
\end{splitdeluxetable*}

\section{Methods of Simulations} \label{sec:methods}

We use \deleted{the MESA }version r-10398 \added{of 1-D stellar evolution code Modules for Experiments in Stellar Astrophysics \citep[MESA,][]{Paxton:2011aa,Paxton:2013aa,Paxton:2015aa,Paxton:2018aa}} to explore the evolution of the presumed progenitor of SN 2013ej from the ZAMS stage through to the core-collapse (CC) stage.\deleted{  An detailed list of physical parameters used for one of our MESA simulations is provided in appendix A (For review purpose only).} All the MESA inlists will be made available publicly at MESA market place\footnote{cococubed.asu.edu/mesa$\_$market/} (subject to acceptance of the manuscript). In the following subsections, we discuss the choices of a few important parameters that affect the results discussed in this paper.

\subsection{Initial Mass and Abundances}

As discussed in the introduction, we investigate isolated (single) star for two different cases of ZAMS masses of 13 \msun \ \& 26 \msun . The initial metallicity Z = 0.006 \added{(same as in Paper I)} is used in our simulations to create a pre-MS model. When provided the Z value, MESA sets up the initial helium abundance (Y) to a value equal to 0.24 + 2$\rm \times Z_{initial}$ \citep[refer to equations (1) through (3) of][]{Choi:2016aa}. As a result, for our simulations the initial H abundance (X) was set up to 0.742 with Y = 0.252, since X+Y+Z = 1.  MESA uses \citep{Grevesse:1998aa} abundance values to derive the initial abundances for each of the metals based on the choice of initial Z. The \citet{Grevesse:1998aa} solar values for helium and metal abundances are \ysun \ = 0.2485 and \zsun \ = 0.0169, respectively.

\subsection{Mass and Temporal Resolution}

In our simulations, we \deleted{incorporated}\added{set the values for} the mass and temporal resolution controls \deleted{described in}\added{to that recommended\footnote{\citet{Farmer:2016aa} explored variations due to number of isotopes in a nuclear reaction network and mass resolution in single, non-rotating, solar metallicity, pre-SN MESA models for a range of masses between 15--30 \msun . Their recommendations are based on convergence of various physical quantities in the star, such as various mass locations, central electron fraction, etc.} in} \citet{Farmer:2016aa}. \deleted{We set both the \texttt{mesh$\_$delta$\_$coeff} \& \texttt{mesh$\_$delta$\_$coeff $\_$for$\_$highT} (applied when $\rm T_{center} \geqslant$ \e{3}{9}) equal to 1.0 in the controls inlist, and varied the mass resolution through \texttt{max$\_$dq} by setting a parameter $\rm \Delta M_{max}$ (\textit{dm} in Table \ref{tab:13m_26m_param}, and hereafter) defined as \texttt{max$\_$dq} = $\rm \Delta M_{max}$ /$\rm M_*(\tau)$. \texttt{max$\_$dq} is the maximum fraction of star mass in a cell, which puts a lower limit on the number of cells in the model. Lower the \textit{dm} (and hence, \texttt{max$\_$dq}) higher is the minimum level of mass resolution. We have tried two values of \textit{dm}, namely, 0.007 \msun \ and 0.01 \msun \ in our simulations for both the stellar masses.}\added{We vary the mass resolution by restricting maximum fraction of star's mass in a cell, \texttt{max$\_$dq}. This is achieved by setting minimum mass resolution $\rm \Delta M_{max}$ or \texttt{dm}, which is defined as \texttt{max$\_$dq} = $\rm \Delta M_{max}$ /$\rm M_*(\tau)$ \citep{Farmer:2016aa}. \added{Adequate mass resolution is required to achieve successful convergence of stellar structure quantities between consequent mass cells. However, very high number of cells require excessive computational resources.} \citeauthor{Farmer:2016aa} recommend \texttt{dm} of 0.01 \msun \ for convergence of various quantities. We used this value \added{in our models. We also tested} \deleted{and }an even finer resolution of 0.007 \msun \deleted{ where it was warranted in our simulations}.} \added{As discussed later in our results section \ref{sec:results}, all our models with both these \texttt{dm} values are sufficiently resolved.}

The time-step between consecutive models in a simulation is generally controlled by \texttt{varcontrol$\_$target}. \added{This is achieved by modulating the magnitude of allowed changes in the stellar variables}. We set it to its MESA default value of \powten{-4}. \deleted{The parameter \texttt{dX$\_$nuc$\_$drop$\_$limit} restricts maximum allowed change in mass fraction above \texttt{dX$\_$ nuc$\_$drop$\_$min$\_$ X$\_$limit}. We chose \texttt{dX$\_$nuc$\_$drop$\_$limit} = \powten{-3}, \texttt{dX$\_$nuc$\_$drop$\_$limit $\_$at$\_$high$\_$T} = \powten{-4}, and \texttt{dX$\_$ nuc$\_$drop$\_$min$\_$X$\_$ limit} = \powten{-3} for evolution from pre-MS to core Si-burning, then onwards we change these limits to \e{5}{-2}, \e{5}{-3}, and \e{5}{-2}, respectively, to allow larger time-steps. In addition, we implemented \texttt{delta$\_$lg$\_$XH$\_$cntr$\_$limit}, \texttt{delta$\_$lg$\_$XHe$\_$cntr $\_$limit}, etc. for all the major nuclear fuels, 
to control time-steps when one of these fuels is depleted in the core. We set these limit to 0.01, and the corresponding hard limit to 0.02 for all of the above species. These limits were applied below a threshold mass fraction at \powten{-6}.}

\added{\citet{Farmer:2016aa} note that the parameters such as \texttt{delta$\_$lg$\_$XH$\_$cntr$\_$limit}, \texttt{delta$\_$lg$\_$XHe$\_$cntr $\_$limit}, etc. offer another useful way of time-step control at critical stages of nuclear burning when one of the major nuclear fuel is depleted in the core such as the TAMS.} This set of limits ensure the convergence of mass shell locations, smoother transition in the HRD at the ``Henyey Hook" \citep{Kippenhahn:2012aa}, and smoother trajectories in the central temperature-density ($\rm T_c-\rho_c$) plane. For a few of our model simulations we had to relax some of these limits at steps where we encountered convergence problems (refer to Table \ref{tab:13m_26m_param} for details). \added{The parameter \texttt{dX$\_$nuc$\_$drop$\_$limit} restricts maximum allowed change in mass fractions between the time-steps when the mass fraction is larger than \texttt{dX$\_$nuc$\_$drop$\_$min$\_$limit}. We set these values similar to that in \citet{Farmer:2016aa}.}

\subsection{Opacities and Equation of State}
We have used Type II opacity tables \citep{Iglesias:1996aa} that take into account varying C \& O abundances \added{during He burning and later stages of evolution,} beyond that accounted for by Z. \deleted{These are particularly useful during He burning and later stages of evolution. }MESA uses Type I tables \citep{Cassisi:2007aa} instead of Type II tables where metallicity is not significantly higher than \texttt{Zbase}\deleted{, a value set by the user. This is because Type I tables cover a wider range of X and have a higher resolution in Z for each X \citep[for more details see][section 4.3]{Paxton:2011aa}. While using Type I opacities \texttt{Zbase} is used instead of actual metallicity}. We set \texttt{Zbase} to the same value as initial Z = 0.006. MESA uses HELM EOS to account for an important opacity enhancement during pair production because of increasing number of electrons and positrons per baryon, at late stages of nuclear burning.

MESA uses equation of states tables based on OPAL tables \citep{Rogers:2002aa}, and at lower densities and temperatures SCVH tables \citep{Saumon:1995aa} with a smooth transition between these tables in the overlapping region defined by MESA.

\subsection{Nuclear Reaction Rates and Networks}

MESA mainly uses NACRE \citep{Angulo:1999aa} rates for thermonuclear reactions with updated triple-$\alpha$ \citep{Fynbo:2005aa} and  \iso{12}{C}($\alpha$,$\gamma$)\iso{16}{O} rates \citep{Kunz:2002aa} among others. \citep[For more details see,][section 4.4]{Paxton:2011aa}. 

In our model simulations\added{, we used \textit{softwired} 79 isotopes network (\texttt{mesa$\_$79.net}, \citeauthor{Farmer:2016aa} \citeyear{Farmer:2016aa}) to optimize between the computational time and convergence of various values at CC stage. For comparison, we present a few simulations with a smaller \textit{hardwired} 22 isotopes network \texttt{approx21$\_$cr60$\_$plus$\_$co56.net} and a single simulation with a \textit{softwired} 45 isotopes network (\texttt{mesa$\_$45.net}) upto O-depletion and 204 isotopes network (\texttt{si$\_$burn.net}, \citeauthor{Renzo:2017aa} \citeyear{Renzo:2017aa}) thereafter until CC. The so called ``hardwired" networks have the predetermined pathway for each reaction, while the ``softwired" networks link all allowed pathways between the isotopes specified in the network. In addition to these simulations, we also present a simulation each for the two ZAMS masses where we used a slightly modified version of the ``test suite" for a pre-CCSN star that accompanies MESA distribution. The test suite uses a 21 isotopes network \texttt{approx21$\_$cr56.net}.
}
\deleted{ that use a 21 or 22 isotopes network, we used the ``extended network" option in MESA. With this option, MESA uses a ``hardwired" eight isotopes network \texttt{basic.net} during hydrogen and helium burning, as it agrees well with a larger networks during these burning stages \citep{Paxton:2011aa}. 
During core carbon burning stage, MESA switches to \texttt{co$\_$burn.net}, which adds the reactions relevant to C/O burning along with $\alpha$ link to \iso{28}{Si} to the \textit{basic} network. For the advanced stages of nuclear burning MESA switches to so called ``advanced" network. At this stage we used a 21 isotopes network \texttt{approx21$\_$cr56.net} in the modified test suite simulations or 22 isotopes network \texttt{approx21$\_$cr60$\_$plus$\_$co56.net} in some simulations listed in Table \ref{tab:13m_26m_param}. The approximate networks are so called ``hardwired" networks where the pathway for each reaction is predetermined. }

\deleted{
The ``softwired" networks, on the other hand, link all allowed pathways between the isotopes specified in the network. For a couple of our simulations presented here, we used such a \textit{softwired} network of 79 isotopes (\texttt{mesa$\_$79.net}, \citeauthor{Farmer:2016aa} \citeyear{Farmer:2016aa}) from ZAMS through CC. We have also presented here one particular simulation where we used a 45 isotopes network (\texttt{mesa$\_$45.net}) upto O-depletion and 204 isotopes network (\texttt{si$\_$burn.net}, \citeauthor{Renzo:2017aa} \citeyear{Renzo:2017aa}) thereafter until CC. 
}

\subsection{Mixing, Diffusion, and Overshoot}

MESA uses standard mixing length theory (MLT) of convection \citep[][chap.14]{Cox:1968aa}. 
We used the Ledoux criterion in our simulations, instead of the Schwarzchild criterion to determine the convective boundaries \citep{Paxton:2013aa}. We used $\rm \alpha_{MLT} =$ 2\added{, which is an intermediate value between the values of 1.6 and 2.2 inferred from comparison of observations with stellar evolution models in the literature \citep[see section 2.3 of][and references therein]{Farmer:2016aa}.} Here the mixing length equals $\rm \alpha_{MLT}$ times the local pressure scale height, $\rm \lambda_P = P/g\rho$. We also enabled two time-dependent diffusive mixing processes, semiconvection and convective overshooting, in our simulations. Semiconvection refers to mixing in regions unstable to Schwarzchild criterion but stable to Ledoux, $\rm \nabla_{ad} < \nabla_{T} < \nabla_{L}$. Semiconvection only applies when Ledoux criterion is used in MESA. We set the dimensionless efficiency parameter, $\alpha_{SC} =$ 0.1 for efficient mixing \added{\citep[same as in][]{Das:2017aa}}. MESA's treatment for overshoot mixing is described in our Paper I \added{in more detail}. \added{The parameter $f_0$ controls the degree of mixing in the overshoot region, whereas the parameter $f$ determines the extent of the overshoot. We have only applied the overshoot for H-burning and non-burning cores and shells.} For the simulations presented in this paper, we keep the \added{overshoot parameter,} $f$ value fixed at 0.025\deleted{ and vary other parameters for the ease of comparison. We try different values of $f_0$ in our simulations to compare the effects this parameter has on the structure and evolution of the progenitor stars. The values of $f_0$ used in our simulations are listed in Table \ref{tab:13m_26m_param}. }

\begin{figure*}[htb!]
\gridline{\fig{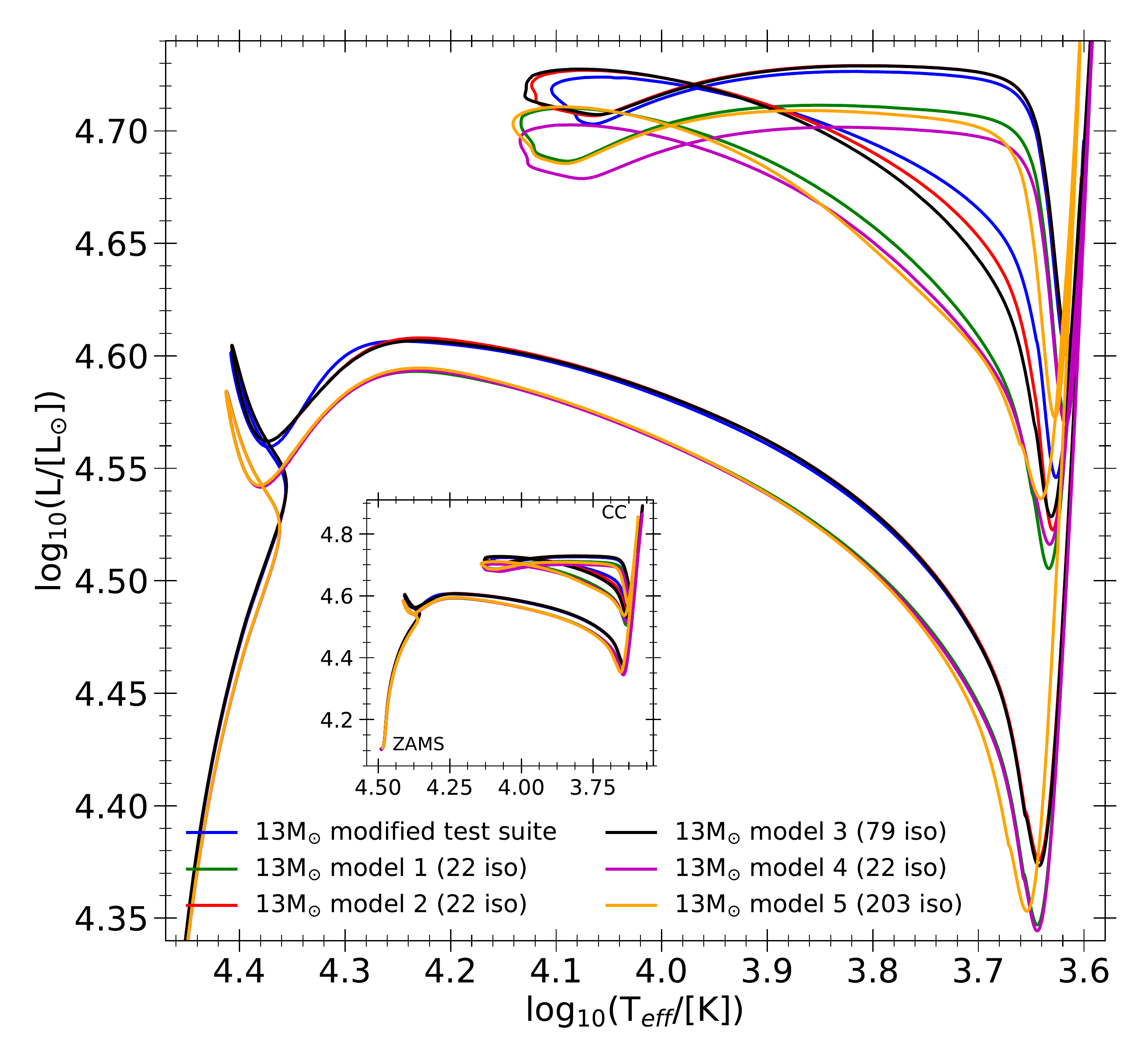}{0.45\textwidth}{(a)}
			  \fig{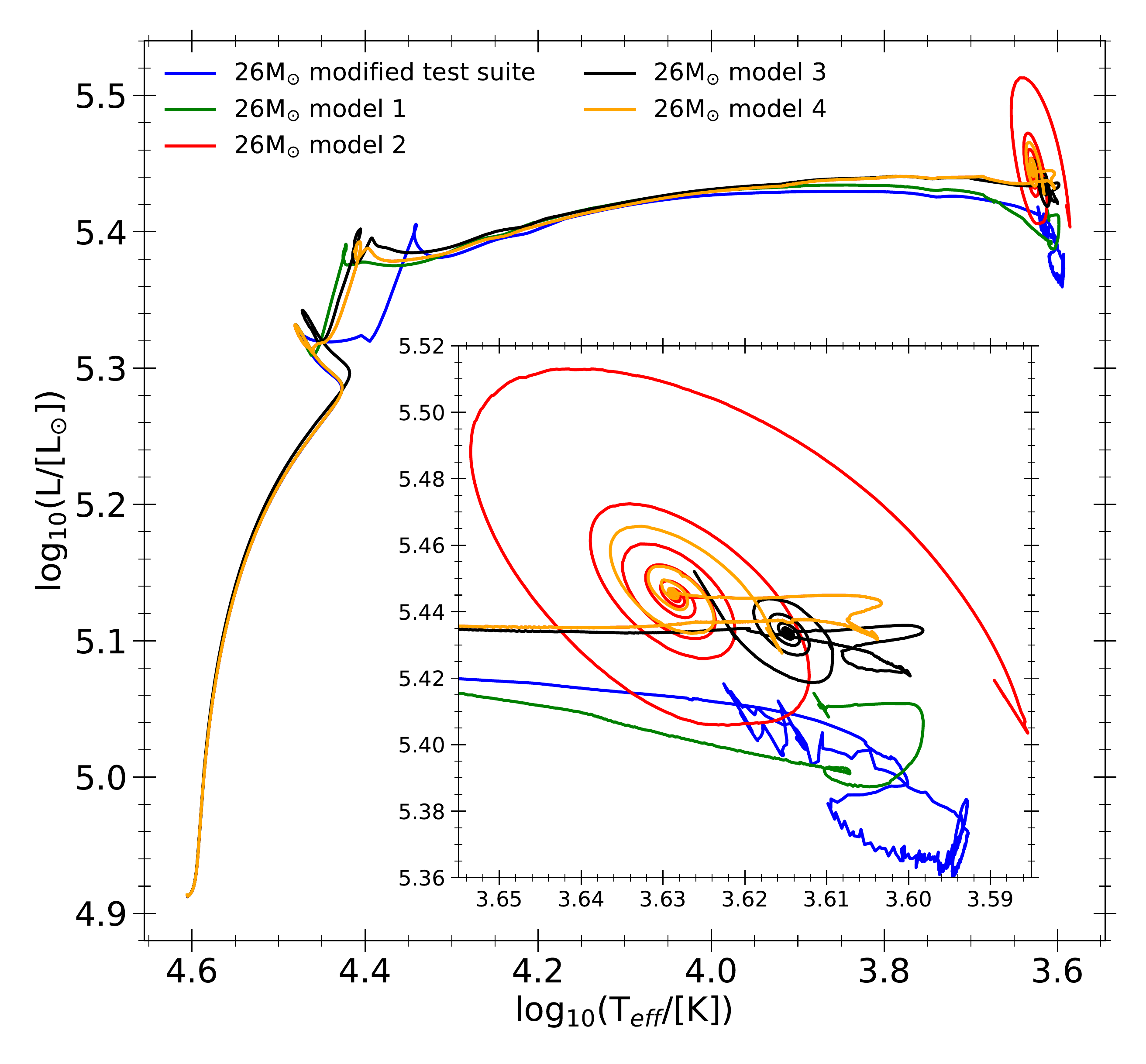}{0.45\textwidth}{(b)}}
\caption{The HRD comparing the evolution for different M$_{ZAMS}$ = 13 \msun \ (a) \& 26 \msun \ (b) simulations with initial Z=0.006 presented in this paper (see Table \ref{tab:13m_26m_param}). The inset in panel (a) shows the whole evolution from ZAMS to CC. 
The 26 \msun \ models with lower mass resolution (\textit{dm} = 0.01 \msun) exhibit spirals (inset in panel (b)) in post Ne-ignition evolutionary track. These spirals go away with a higher mass resolution (\textit{dm} = 0.007 \msun , 26 \msun \ model 1), but it is unaffected by a finer temporal resolution (max change in \textit{dt} = 1.15, model not shown here). 
\label{fig:13M_26M_HR}}
\end{figure*}

\subsection{Mass Loss by Stellar Winds}
\deleted{In our simulations, e implement mass loss schemes for the ``hot" and the ``cool" phases of evolution. }For the ``hot" phase\added{ of evolution}
, we used the ``Vink" scheme \citep{Vink:2001aa} with $\eta_{Vink}$ = 1.0 (\texttt{Vink$\_$scaling$\_$ factor}) and for the ``cool" phase, 
we used the ``Dutch" wind scheme for both the AGB and the RGB phases, with $\eta_{Dutch}$ = 0.5 (\texttt{Dutch$\_$scaling$\_$factor}) in the simulations presented here. \added{This particular combination is used to confine the mass loss rate to moderate levels during the RSG phase. We note that the same combination was used by \citet{Das:2017aa}\footnote{Note that the value of $\eta_{Dutch}$ was erroneously quoted as 1.0 in Appendix A of \citet{Das:2017aa}. Their table 1 and Figures describe the models that use $\eta_{Dutch}$ = 0.5, not 1.0 as quoted.}.}
\deleted{ For massive stars, the ``Dutch" scheme combines results from several authors, mostly from Netherlands.} The particular combination used \added{for the ``Dutch" scheme} in MESA is based on \citet{Glebbeek:2009aa}. 
The MESA mass loss rates are switched between the rates of \citet{Vink:2001aa,Vink:2000aa} to that of \citet{de-Jager:1988aa} at temperatures below 10,000 K. 

\subsection{Stopping criterion}
The model is stopped very close to the collapse of the star when the infall velocity at any location in the interior of the star reaches the \texttt{fe$\_$core$\_$infall$\_$limit} = \powten{8} $\rm cm~s^{-1}$. We turned on the velocity variable `v' defined at cell boundaries, by setting \texttt{v$\_$flag = .true.} when the central electron fraction $\rm Y_e$ (electrons per baryon, $\rm \bar{Z}/\bar{A}$)  drops below 0.47 (\texttt{center$\_$ye$\_$limit$\_$for$\_$v$\_$flag}). This enables MESA to compute hydrodynamic radial velocity, which is used for evaluating the stopping criterion.


\begin{figure*}[htb!]
\gridline{\fig{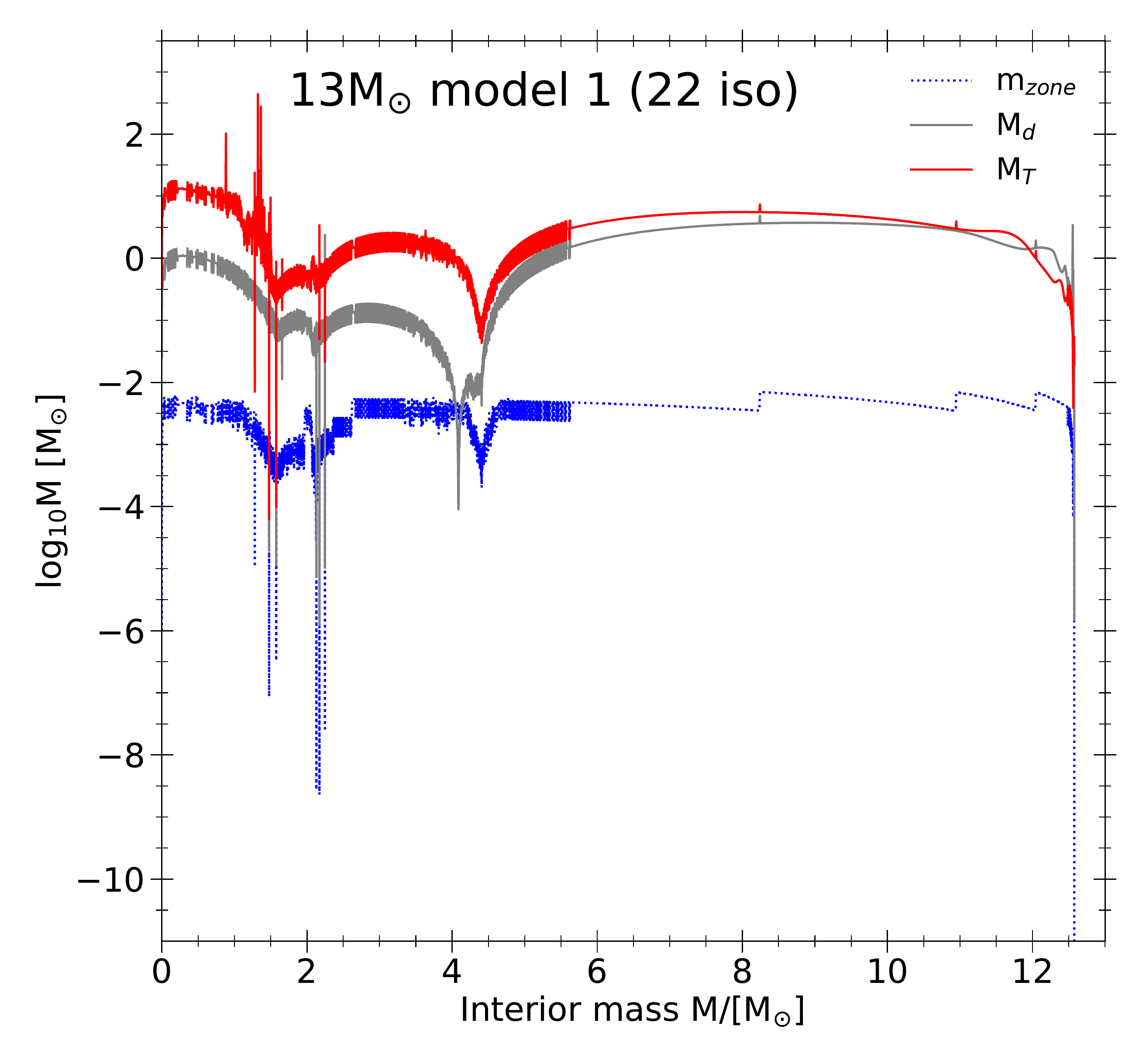}{0.33\textwidth}{(a)}
			    \fig{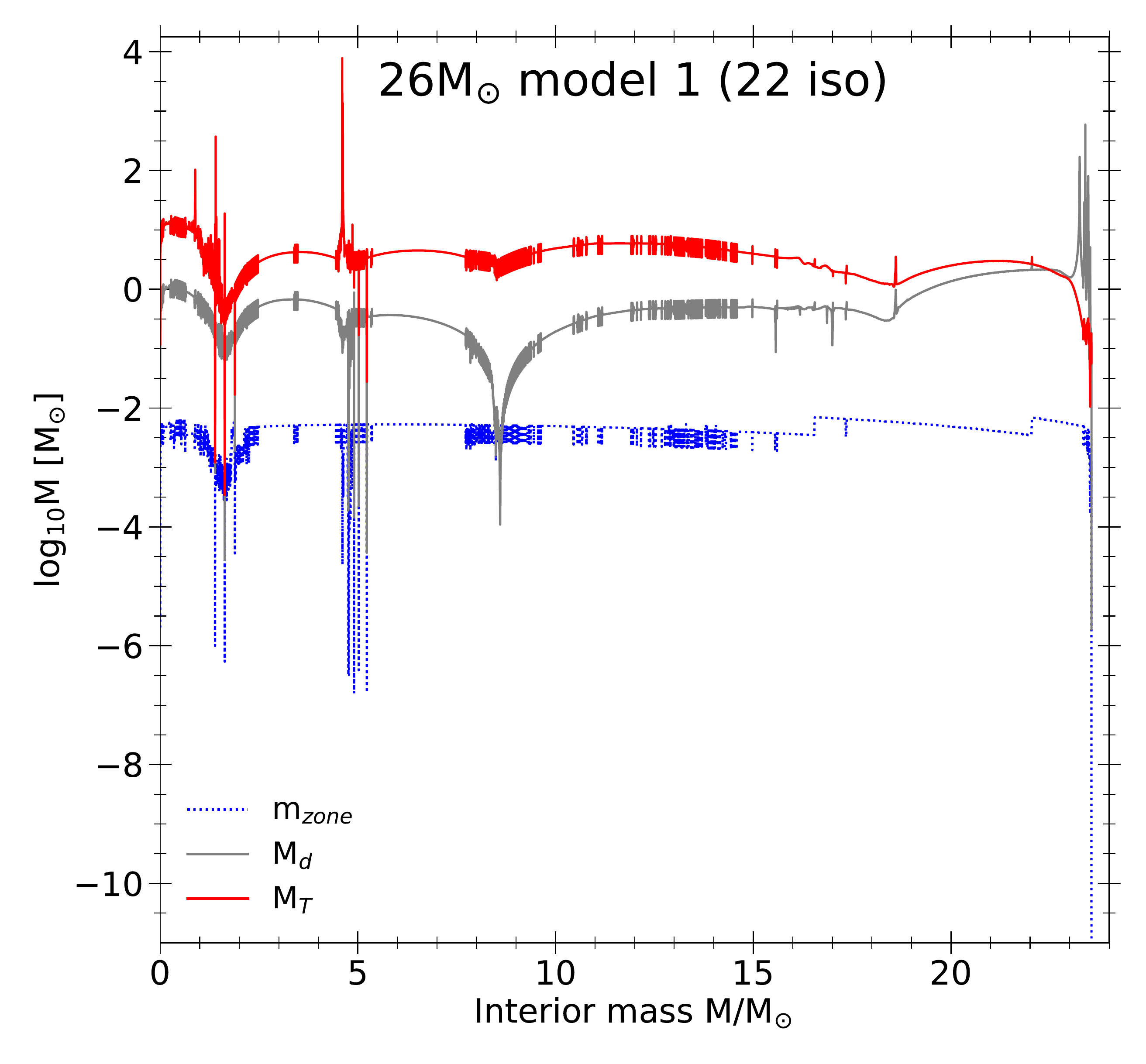}{0.33\textwidth}{(b)}
				\fig{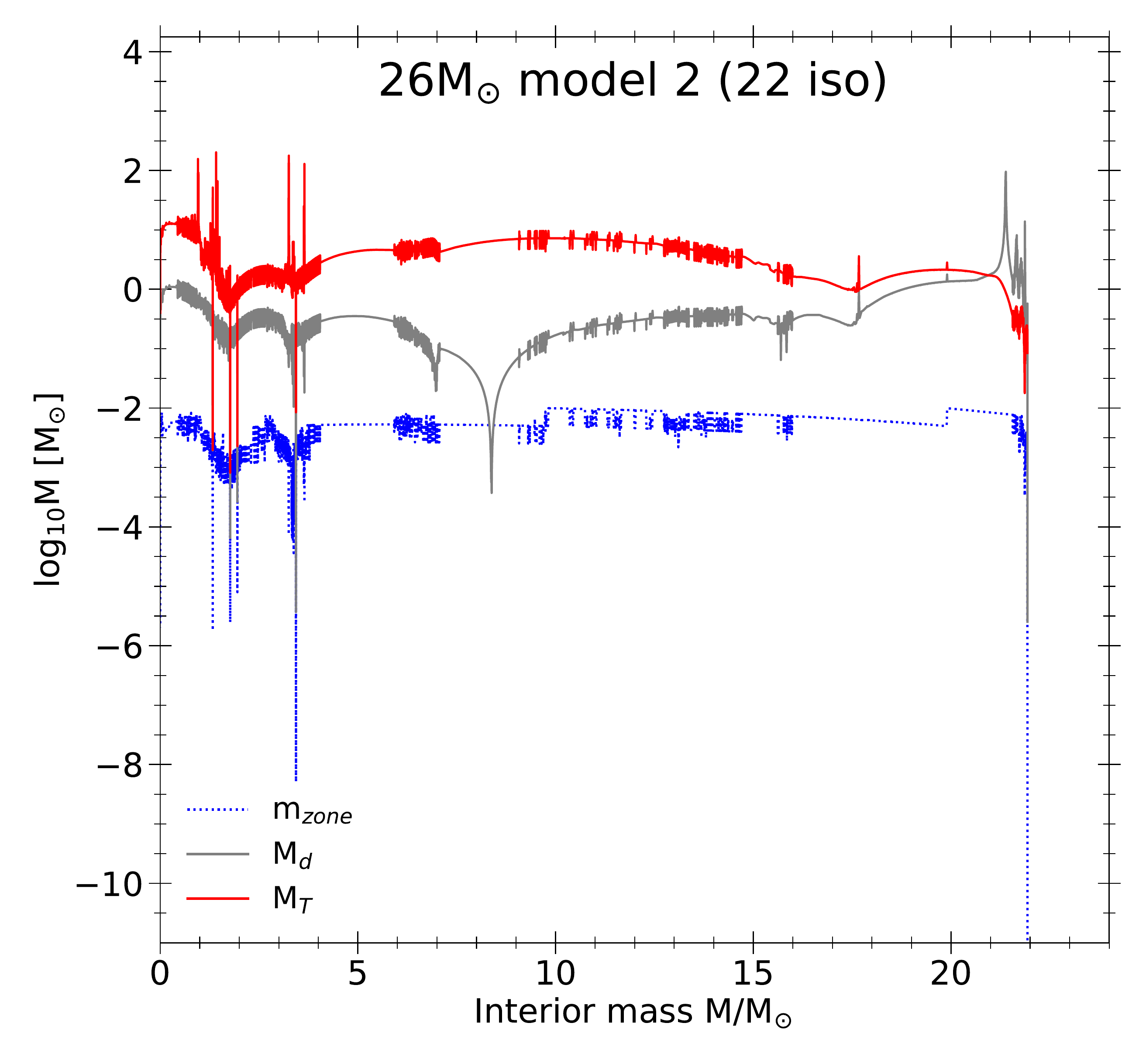}{0.33\textwidth}{(c)}
				}
\caption{The density (grey) and temperature (red) scale heights along with the mass resolutions (zoning, blue dotted line) for the M$_{ZAMS}$= 13 \msun \ model (a) \& 26 \msun \ models (b,c) with Z = 0.006 at CC. The scale height M$_x$ = (d ln x/dm)$^{-1}$ \citep{Sukhbold:2018aa} for a quantity `x' to mass resolution ratio is an indicator of zoning efficiency. Panel (a) shows the 13 \msun \ model with \textit{dm} = 0.007 \msun , and panels (b) \& (c) show 26 \msun \ models 1 \& 2 with  \textit{dm} =  0.007 \msun \ and 0.01 \msun , respectively. The fluctuations in density and the pressure scale heights is over-resolved in our simulations, which is apparent by the 2 dex fine zoning at most places. 
\label{fig:13M_26M_mass_res}}
\end{figure*}

\section{Results} \label{sec:results}

\deleted{The choices for input parameters for our 13 \msun \ \& 26 \msun \ simulations are identical except for one or two key parameters that are varied between the simulations as listed in Table \ref{tab:13m_26m_param}, respectively, and described in section \ref{sec:methods}. Here, we discuss the effects of these variations on various surface and central properties of the star. }

\begin{deluxetable*}{ccccccccccc}[htb!]
\tablecaption{Magnitudes and Dust Extinction for  \Mzams \ = 26 \msun, Z = 0.006 model 1
\label{tab:dust_ext}}
\tablehead{
\colhead{HST} & \colhead{Johnson} & \colhead{Synthetic} & 
\colhead{MESA} &  \colhead{Min.} & 
\multicolumn{5}{c}{Calculated $A(\nu)$ for Graphite} 
& \colhead{Calculated}  \\ [-2ex]
\colhead{Observed} & \colhead{Band} & \colhead{Observed} & 
\colhead{Calculated} & \colhead{$A(\nu)$} & \colhead{minimum} &
\multicolumn{2}{c}{Heat Balance} & \multicolumn{2}{c}{Adiabatic Cooling} & 
\colhead{$A(\nu)$ for} \\ [-2ex]
\colhead{Magnitude} & \colhead{}  & \colhead{Magnitude\tablenotemark{a}} & \colhead{Magnitude} & \colhead{required} &  
\colhead{grain} & \colhead{$\rm T_{d,max}$} & \colhead{$\rm T_{d,max}$} & 
\colhead{$\rm T_{d,max}$} & \colhead{$\rm T_{d,max}$} 
& {Silicates\tablenotemark{c}} \\ [-2ex]
\colhead{} & \colhead{} & \colhead{($\rm m_{obs}$)} & \colhead {($\rm M_{MESA}+ \mu_d$)\tablenotemark{b}} & \colhead{} & 
\colhead{\AA} & \colhead{1500} & \colhead{2000} & \colhead{1500} & 
\colhead{2000} & \colhead{}
}
\colnumbers
\startdata
\multicolumn{11}{c}{\boldmath $(\rm \tau_{CC} - \tau) \approx$ \textbf{10 yrs}, \boldmath$\rm R_{star} = 1017.73~ R_{\odot}$} \\
\hline
\multicolumn{1}{c}{\multirow{2}{*}{25.16$\pm$0.09}} & \multicolumn{1}{c}{\multirow{2}{*}{V}} & \multicolumn{1}{c}{\multirow{2}{*}{24.99$\pm$0.13}} & \multicolumn{1}{c}{\multirow{2}{*}{21.91}} & \multicolumn{1}{c}{\multirow{2}{*}{3.08}} & 800 & 1.35 & 3.89 & 4.51 & 5.28 & \multicolumn{1}{c}{\multirow{2}{*}{0.74}} \\
\cline{6-10}
& & & & & 1500 & 1.13 & 3.27 & 3.80 & 4.44 \\
\hline
\multicolumn{1}{c}{\multirow{2}{*}{22.66$\pm$0.03}} & \multicolumn{1}{c}{\multirow{2}{*}{I}} & \multicolumn{1}{c}{\multirow{2}{*}{22.70$\pm$0.09}} & \multicolumn{1}{c}{\multirow{2}{*}{20.40}} & \multicolumn{1}{c}{\multirow{2}{*}{2.30}} & 800 &  0.65 & 1.86 & 2.16 & 2.53 & \multicolumn{1}{c}{\multirow{2}{*}{0.36}} \\
\cline{6-10}
& & & & & 1500 & 0.54 & 1.57 & 1.82 & 2.13 \\
\hline
\multicolumn{11}{c}{\boldmath $(\rm \tau_{CC} - \tau) \approx$ \textbf{8 yrs}, \boldmath$\rm R_{star} = 1017.69~ R_{\odot}$} \\
\hline
\multicolumn{1}{c}{\multirow{2}{*}{24.84$\pm$0.05}} & \multicolumn{1}{c}{\multirow{2}{*}{V}} & \multicolumn{1}{c}{\multirow{2}{*}{24.69$\pm$0.08}} & \multicolumn{1}{c}{\multirow{2}{*}{21.91}} & \multicolumn{1}{c}{\multirow{2}{*}{2.78}} & 800 & 1.35 & 3.90 & 4.51 & 5.32 & \multicolumn{1}{c}{\multirow{2}{*}{0.74}} \\
\cline{6-10}
& & & & & 1500 & 1.14 & 3.28 & 3.79 & 4.47\\
\hline
\multicolumn{1}{c}{\multirow{2}{*}{22.66$\pm$0.03}} & \multicolumn{1}{c}{\multirow{2}{*}{I}} &  \multicolumn{1}{c}{\multirow{2}{*}{22.69$\pm$0.08}} & \multicolumn{1}{c}{\multirow{2}{*}{20.40}} & \multicolumn{1}{c}{\multirow{2}{*}{2.29}} & 800 & 0.65 & 1.87 & 2.16 & 2.55 & \multicolumn{1}{c}{\multirow{2}{*}{0.36}} \\
\cline{6-10}
& & & & & 1500 & 0.54 & 1.57 & 1.82 & 2.14  \\
\enddata
\tablecomments{The calculated (absolute) magnitudes from our MESA simulation are listed along with the observed HST WFC magnitudes in F555W \& F814W bands from \citet{Fraser:2014aa}. The dust extinction values calculated using the formalism explained in Appendix \ref{app:dust_ext} are listed here.}
\tablenotetext{a}{The synthetic magnitudes in V \& I bands is calculated from HST observations using the algorithm explained in \citet{Sirianni:2005aa}.}
\tablenotetext{b}{A distance modulus $\mu_d$ = 29.77$^{+0.09}_{-0.1}$ is calculated considering the distance of M74 to be 9$^{+0.4}_{-0.6}$ Mpc.}
\tablenotetext{c}{A single value of $T_{d,max}$ = 1500 K was used to find $R_{min}$ for silicates, as silicates do not withstand a higher temperature.}

\end{deluxetable*}

\subsection{HR diagram \& mass resolution}

\added{Fig. \ref{fig:13M_26M_HR} shows the evolutionary tracks in the HRD for the 13 \msun \ star in panel (a) and 26 \msun star in panel (b).}
\deleted{As seen in Fig. \ref{fig:13M_26M_HR} panel (a), the evolutionary tracks in the HRD for our 13 \msun \ simulations are close to each other}\added{Various 13 \msun models closely track each other throughout the evolution of the star with minor differences stemming from differences in choices of parameters.} \deleted{However, we can see them grouped based on the choices of $f_0$ and mass loss $\eta$ factor. 13 \msun \ models 2 (red) \& 3 (black) with same f/f$_0$ values of 0.025/0.005 are in a good agreement with each other with slight differences due to choice of network. Model 2 with 22 isotopes stays a little redder than the 79 isotope model 3. Model 1 (green), 4 (purple), \& 5 (orange) have f = 0.025 and $f_0$ = 0.050 \citep[as in][]{Das:2017aa}. These models typically remain less luminous throughout than Models 2 \& 3 with a lower $f_0$ value. Models 1 \& 5 differ in choice of network size, while model 4 has higher mass loss rate and lower \textit{dm} value. Model 4 stays a little less luminous and redder than models 1 \& 5.}\added{Model 4 with $\eta_{Dutch}$ = 1.0 remains less luminous than other models with $\eta_{Dutch}$ = 0.5. The variations in the models with same $\eta_{Dutch}$ which are due to differences in the choices of one or two parameters (see Table \ref{tab:13m_26m_param}) are not significant to distinguish through observations.}

On the other hand, we notice that the effects of choices of parameters are more prominent in the 26 \msun \ model, especially with \textit{dm} which controls the minimum level of mass resolution. \deleted{As seen in Fig. \ref{fig:13M_26M_HR} panel (b), 26 \msun \ models 2 (red), 3 (black), and 4 (orange) that use \textit{dm} = 0.01 \msun, show spirals in their HR tracks starting near C-shell burning towards core Ne-ignition [$(\tau_{CC} - \tau) \sim$ 20 yrs]. The extent of these spirals appear to be smaller for higher isotope network in model 4 than in model 2, and even smaller in model 3 (black) that uses a smaller $f_0$ value.}\added{We see in panel (b) for the HRD that models 2, 3 \& 4 that use a coarser mass resolution (\textit{dm} = 0.01 \msun ) exhibit spirals in their evolutionary tracks around core Ne-ignition. This feature is independent of the choices for other parameters.} The most variation in luminosity takes place in model 2 (red) between 1.3 years and 0.6 years before collapse from log L = 5.4 to 5.5 (see the inset in panel (b))\deleted{ for reference)}. \deleted{This variation in luminosity corresponds to 0.3 mag in V \& I band magnitudes (as reported by MESA)}\added{The corresponding variation in V \& I band magnitudes reported by MESA is 0.3 mag}, which can be easily observed. Unfortunately, \deleted{for the SN 2013ej progenitor there exist}\added{the} observations \added{of the progenitor of SN 2013ej are available} only \added{at} about 8 \& 10 years before its collapse. The variation in magnitudes \deleted{for any of our simulations at this stages is less than 0.05 mag, which is}\added{during 10 to 8 years before collapse in the models above is consistent with the HST observations.}\deleted{ not significant enough to be able to confidently rule out any of the above models. A point to be noted is that}\added{ Note however that} the spirals are not exhibited in model 1 that uses a \deleted{lower}\added{finer mass resolution (}\textit{dm} \deleted{value of}\added{= 0.007 \msun )}\deleted{, which controls the minimum mass resolution (zoning). 
}
The modified test suite models for both the ZAMS masses explored here are presented for comparison purpose only. The test suite models suffer from lack of strict temporal and mass resolution, the effect of which is prominent in the inset of panel (b) for 26 \msun \deleted{where there is a lot of ``noise" in the HR track}.

\deleted{Though it seems that the spirals in the HRD are consequence of coarser resolution, we note that all our 13 \msun \ and 26 \msun \ models that use strict mass resolution controls are adequately resolved (as seen in Fig. \ref{fig:13M_26M_mass_res}). One of the criteria described by}\added{According to} \citet{Sukhbold:2018aa} \added{one of the criteria} for adequate zoning is that the key variables like density and temperature do not vary significantly between consecutive zones.  In Fig. \ref{fig:13M_26M_mass_res}, we plot density and pressure scale heights along with the mass resolution \added{through the interior of the}\deleted{ for our}\added{ 13 \msun \ \& 26 \msun \ stars. The} 13 \msun \ model 3 with \textit{dm} = 0.007 is shown in panel (a) and the 26 \msun \ models 2 \& 3 with \textit{dm} = 0.007 \msun \ and 0.01 \msun \ are shown in panels (b) \& (c), respectively. A scale height for a quantity 'x' is defined as M$_x$ = (d ln x/dm)$^{-1}$, where an absolute value of M$_x$ is considered. The large upward spikes in Fig. \ref{fig:13M_26M_mass_res} represent regions of near constant temperature and density. The discontinuities in mean molecular weight at the edge of convective shells result in quantities being artificially small. The edge of the helium convective shell (mass coordinate $\sim$ 4 \msun \ in 13 \msun \ model and 8 \msun \ in 26 \msun \ model) shows steep gradient where the density changes by a few orders of magnitude, where the zoning is not well resolved\deleted{, even with \textit{dm} = 0.007 \msun }. A finer resolution comes at a cost of increased computational time\deleted{, without any benefits to overall results of the simulations \citep[as noted in][]{Sukhbold:2018aa}}. Other than the edge of He shell, the zoning is 2 dex finer than the scale heights \added{everywhere inside both the stars} indicating that the quantities are over-resolved \deleted{in all of our simulations presented in the figure}\added{for both the \textit{dm} values}.\deleted{ Hence, we may not be able to completely rule out the spirals appearing as numerical artefacts without observational evidence. 
}

\subsection{Dust Extinction for 26 \msun \ progenitor}

Table 1 of \citet{Fraser:2014aa} lists the observed magnitudes from archival pre-explosion images of the progenitor for SN 2013ej. The observations made in Nov 2003 \& June 2005 are available, at about 10 and 8 years before the collapse of the star. We converted the HST observed magnitudes in F555 and F814 bands to Johnson's photometric magnitudes V and I using formalism explained by \citet{Sirianni:2005aa}. To compare the MESA calculated V \& I magnitudes to the observations, we evaluated the dust extinction due to the stellar wind formed by circum-stellar medium (CSM), using the formalism provided in Appendix \ref{app:dust_ext}. These values are listed in Table \ref{tab:dust_ext} along with the HST observed magnitudes and the MESA calculated magnitudes for \added{26 \msun \ fiducial star}\deleted{our simulations}. We notice that the V band magnitude is reduced by about 0.3 mag in $\sim$2 years from 2003 to 2005. Our \added{MESA} calculations do not show such variations for any of our 13 \msun \ or 26 \msun \ simulations between 10 to 8 years before collapse. \added{The dust extinction values calculated based on the stellar wind model do not account for this variation either.}\deleted{ It is possible that this variation is connected to the extinction by the CSM around the star, however, our calculations of extinction values do not account for this variation either.}

We see from Table \ref{tab:dust_ext} that a minimum A$_{\rm V} \approx$ 2.8 -- 3.1 is required to shield our fiducial 26 \msun \ progenitor. Although we see such high numbers among our calculated A$_{\rm V}$ values (columns (6) through (10) in the table),\deleted{ especially with the assumption that adiabatic cooling of gas due to expansion is used to find minimum distance $R_{min}$ at which dust can form. However, the adiabatic cooling of gas is not sufficient to form dust, as there is a strong radiation source close to the gas that will evaporate the dust. So, using heat balance equation to determine $R_{min}$ is more appropriate. We also see high values of A$_{\rm V}$ when we choose $T_{d,max}$ = 2000 K \citep{Fox:2010aa}. A minimum required A$_{\rm V} \approx$ 3 [$\rm m_{V,obs} -(M_{V,MESA} + \mu_d$)]}\added{ it} is significantly higher compared with the value of 0.45$^{+0.04}_{-0.04}$ calculated by \citet{Maund:2017aa} in the host for the progenitor of SN 2013ej through Bayesian analysis of stellar population, using archival pre-SN data. As the host galaxy M74 is observed face-on, most of this extinction would be due to the CSM around the star. For another supernova of Type IIP, SN 2003gd in the same host galaxy M74 as SN 2013ej, \citet{Maund:2017aa} calculates A$_{\rm V}$ = 0.46$^{+0.04}_{-0.03}$.

%


\deleted{The compactness parameter defined
 and discussed in Paper I has been shown in Figures \ref{fig:13M_xi} and \ref{fig:26M_xi} for 13 \msun \ and 26 \msun \ stars. The most notable aspect of these two figures is that for 13 \msun \ star,  the $\xi_{2.5}$ parameter is much smaller than the corresponding parameter for 26 \msun \ star (whose asymptotic values at the CC stage are typically between 0.13 to 0.19). This makes the 13 \msun \ stars much more amenable to explosion according to \citet{OConnor:2011aa} than the 26 \msun \ star. We note however, that the scales of the $\xi_{1.6}$ parameter for both 13 and 26 \msun \ stars are much more similar, though even here the 13 \msun star has a smaller $\xi_{1.6}$.}

\subsection{Central Temperature versus Density}

\begin{figure*} [htb!]
\gridline{\fig{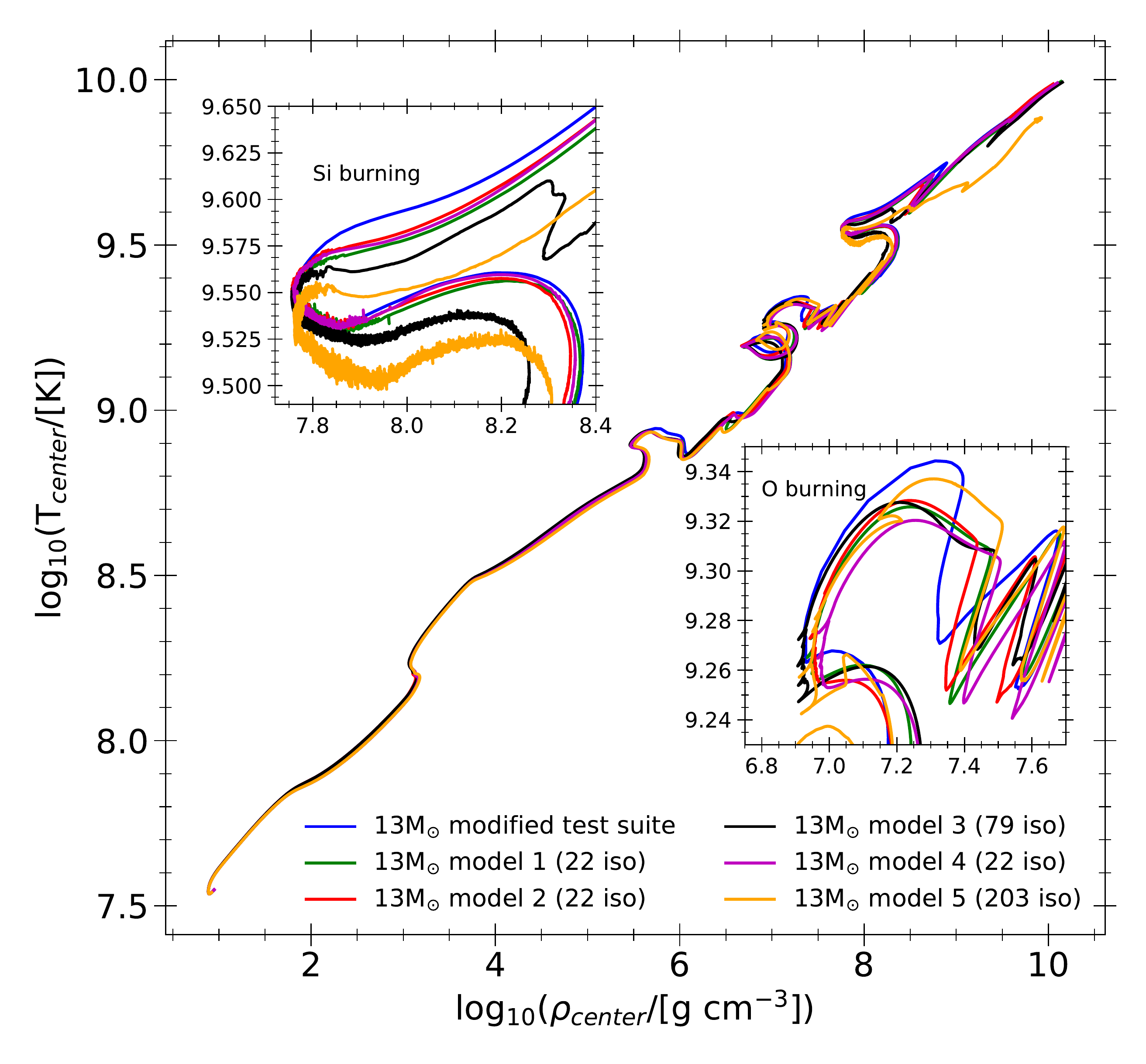}{0.45\textwidth}{(a)}
			  \fig{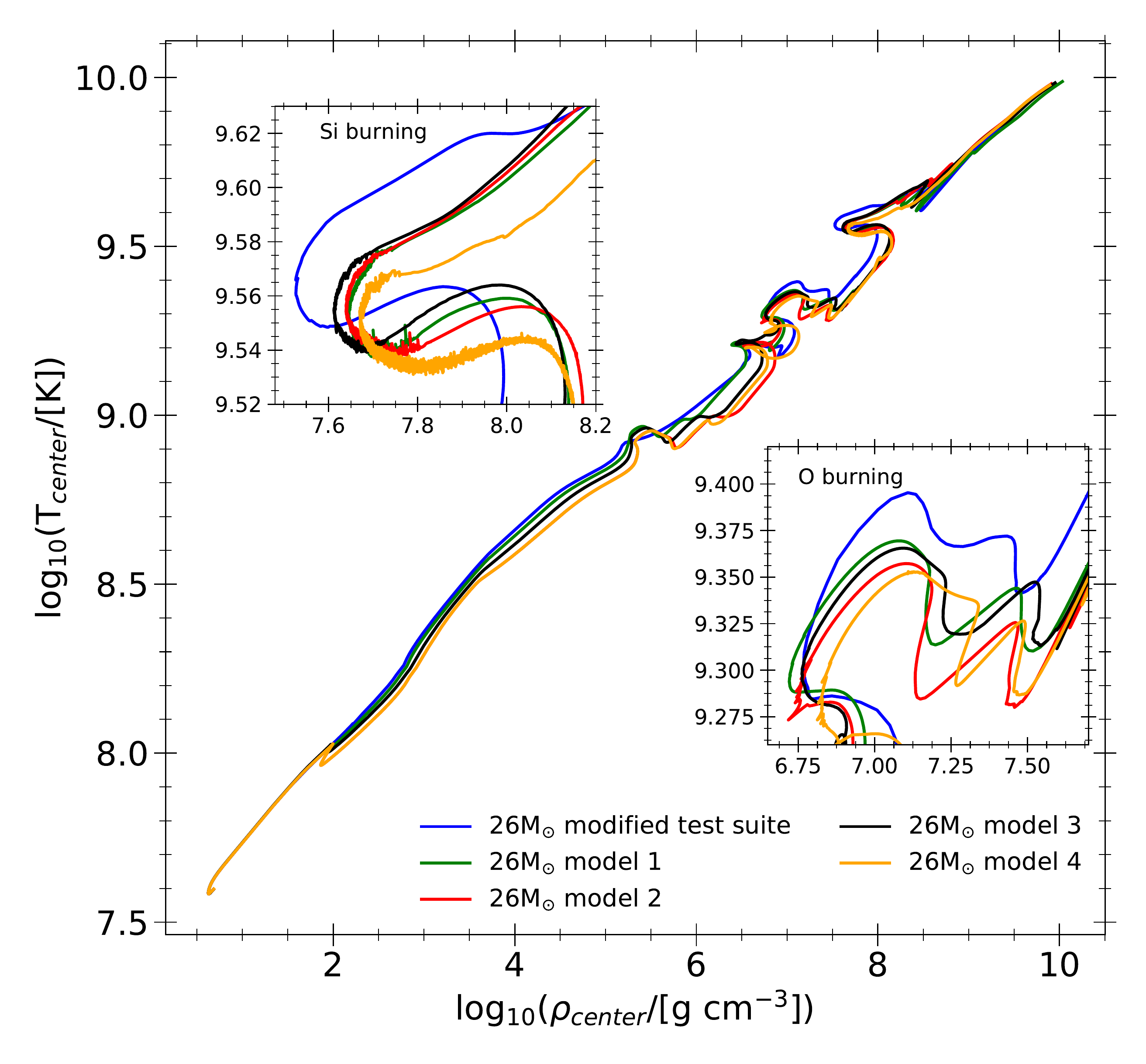}{0.45\textwidth}{(b)}}
\caption{$\rho_c -$ T$_c$ plot comparing the M$_{ZAMS}$= 13 \msun \  and 26 \msun \ models, with Z = 0.006 described in Table \ref{tab:13m_26m_param}. Panels (a) and (b) show the evolution from ZAMS through CC for 13 \msun \ and 26 Sun models respectively, while the two sets of insets in these show the same during O-ignition through Si burning. One can observe that there are wiggles present during O and Si burning stages in the plot, which in this case, are present in almost all of the models. The wiggles during O burning stage disappear in 26 \msun \ model when higher mass resolution (\textit{dm} = 0.007 \msun) is chosen. The wiggles may therefore be numerical artefacts.
\label{fig:13M_26M_rhoc_Tc}}
\end{figure*}

Fig. \ref{fig:13M_26M_rhoc_Tc} shows evolution of central temperature and density from ZAMS through CC. The 13 \msun \ models shown in \deleted{the left} panel (a) are in a good agreement \added{with each other} until about when C-burning starts \added{in the core} ($log_{10}T_{c} \approx$ 8.9 and $log_{10}\rho_{c} \approx$ 5.5). The 26 \msun \ models shown in \deleted{the right} panel (b) diverge after TAMS stage ($log_{10}T_c \approx$ 8 and $log_{10}\rho_c \approx$ 2). We notice that models with smaller network size tend to be hotter and less dense at the center through various burning stages. The insets in both the panels show evolutionary tracks during core O-burning and core Si-burning stages. We note that the tracks in this part \deleted{are very "noisy"}\added{show rapid variations}, more so during core Si-burning, indicating that these phases of nuclear burning are challenging due to very high and nearly balancing reaction rates, as also pointed out in \citet{Renzo:2017aa}. \deleted{In addition, we note that our simulations with neither higher network size nor higher zoning resolution are able to reduce this noise.}

\subsection{The Kippenhahn diagram} 

\begin{figure*}[htb!]
\gridline{\fig{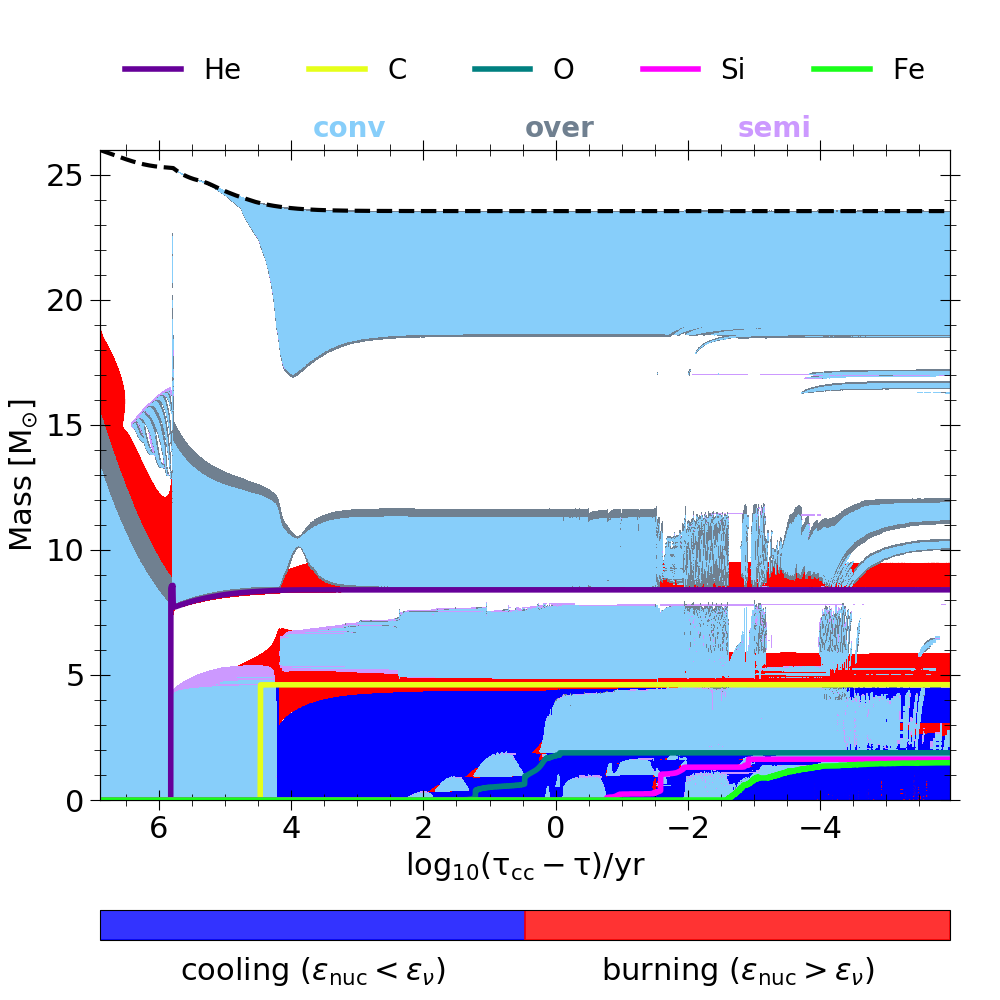}{0.45\textwidth}{(a)}
				\fig{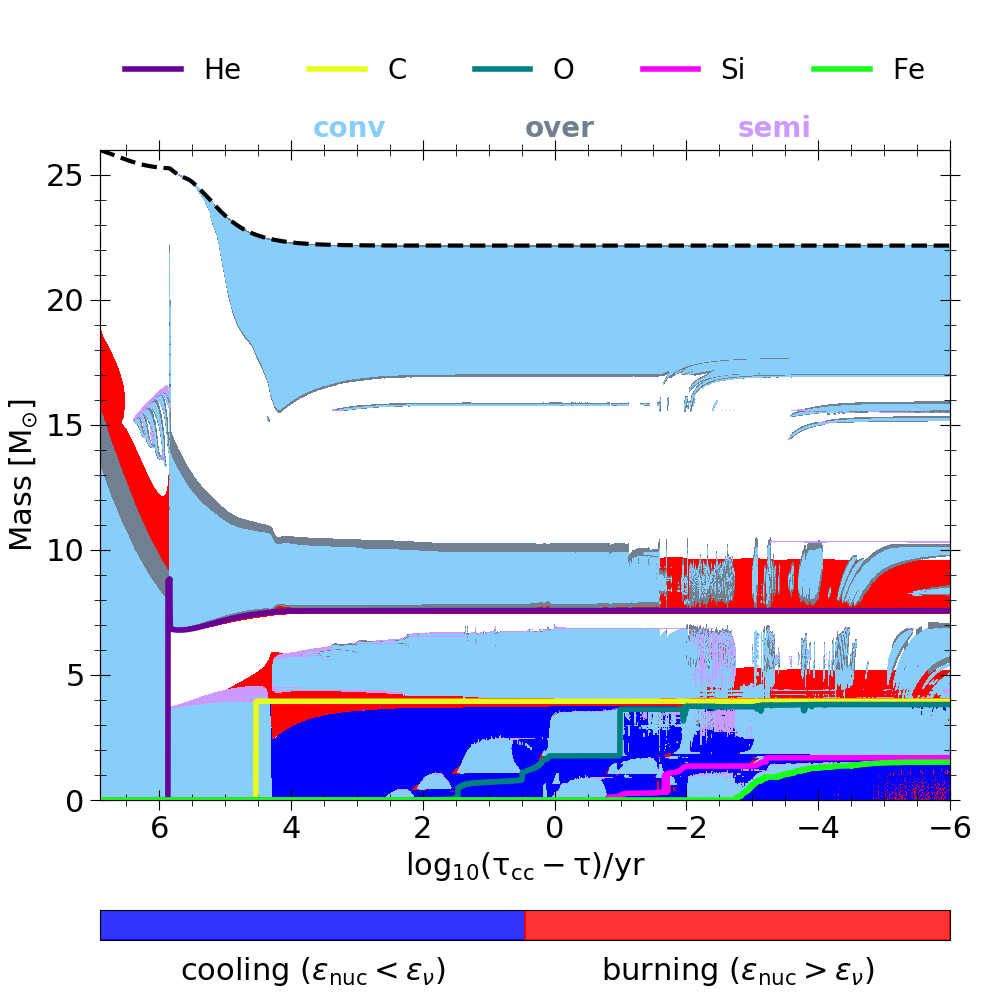}{0.45\textwidth}{(b)}}
\gridline{\fig{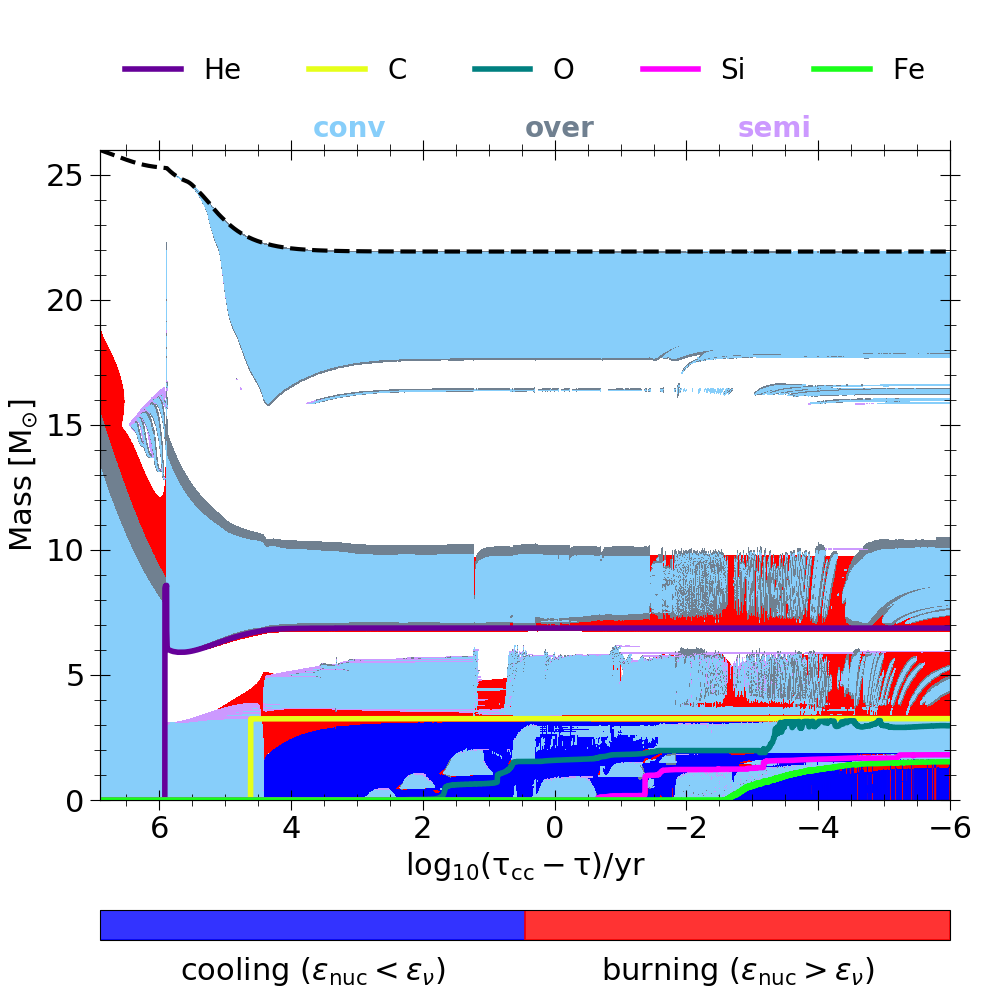}{0.45\textwidth}{(c)}}
\caption{\deleted{Same as Fig. \ref{fig:13M_KH} but for} Kippenhahn (KH) diagram for the M$_{ZAMS}$= 26 \msun \ and Z = 0.006  models described in Table \ref{tab:13m_26m_param}. Panel (a) shows model 1 with \textit{dm} = 0.007 \msun \ and $f_0$ = 0.050, while panel (b) shows model 3  with \textit{dm} = 0.01 \msun \ and $f_0$ = 0.020, both using 22 isotopes network. Panel (c) shows model 4 with \textit{dm}  = 0.01 \msun \ and $f_0$ = 0.050, with 79 isotopes network. 
Various core boundaries reported by MESA are marked in the diagram. By default, the core boundaries are determined by the point at which the mass fraction of the previous isotope decreases below a threshold value of 0.01 and where the same for the current isotope is above 0.1.
\label{fig:26M_KH}}
\end{figure*}

The Kippenhahn \added{(KH)} diagram is an useful representation of the succession of various convective burning zones inside the star throughout its evolution.\deleted{ In Fig. \ref{fig:13M_KH}, we show 13 \msun \ models 2 \& 3 that differ in choices of the overshoot parameter $f_0$ and network size but have the same mass resolution \textit{dm} = 0.007 \msun \ and the overshoot parameter $f$ = 0.025. We have only applied the overshoot for H-burning and non-burning cores and shells. 
We notice that the radial extent of the overshoot (in grey) during the main-sequence phase (up to a million years before collapse) is same for both the models with different $f_0$ values (unlike the other overshoot parameter $f$, as discussed in our Paper I.) The parameter $f_0$ controls the degree of mixing in the overshoot region, whereas the parameter $f$ determines the extent of the overshoot. The core boundaries defined by MESA are also plotted for various major nuclear fuels. We see that the overall structure of the star (mass location of various burning zones, duration of burning, total star mass, etc.) is less affected by choices $f_0$ and network size. However, one can see some differences in the finer structures of the burning (red) and convective (sky blue) regions. Model 1 has three well defined C-burning shells after core C burning, while in model 3 the third C-shell (at $\sim$ 10 years to CC) appears to be sporadic. This difference is due to the choice of the overshoot parameter $f_0$.}
\added{We have discussed the KH diagram for 13 \msun \ star in Paper I. In }
Fig. \ref{fig:26M_KH}\deleted{, shows}\added{ we compare} 26 \msun \ model 1 with \textit{dm} = 0.007 \msun \ and $f_0$ = 0.050, model 3 with \textit{dm} = 0.01 \msun \ and $f_0$ = 0.020, and model 4 with \textit{dm} = 0.01 \msun \ and $f_0$ = 0.050. Model 1 \& 3 both use 22 isotopes network while model 4 uses 79 isotopes network.\deleted{ As seen from this figure the models with same mass resolution have more similarities in the internal structure throughout the history of the star, and is less dependent on the choice of network size or $f_0$, as in the case of 13 \msun \ models.} Note that 26 \msun \ models 3 \& 4 \added{(panels (b) and (c) in Fig. \ref{fig:26M_KH})} both exhibiting spirals in the HRD, have a distinct intermediate convective zone (ICZ) at the bottom of the convective hydrogen shell (at about 16 \msun) starting at about a few thousand years before the collapse, albeit the ICZ being very limited in mass extent\deleted{, which}\added{. This ICZ} is missing in model 1 \added{(panel (a) in Fig. \ref{fig:26M_KH})}.\deleted{ Model 1 with \textit{dm} = 0.007 \msun \ has a continuous C-burning shell around the time neon is ignited in the core at about 1 year before CC, while in models 3 \& 4 the same C-shell sporadically turned off and on again. Note that this difference, in contrast to the case of 13 \msun , is due to the finer mass resolution in model 1. Unlike $f_0$,} The mass resolution \added{and the network size} affect the internal structure of the star. The various core boundaries in the \deleted{higher}\added{coarser mass resolution} (\textit{dm} = 0.01 \msun ) models 1 \& 4 are systematically lower than those in the \added{finer mass resolution} (\textit{dm} = 0.007 \msun ) model 1. \added{The star is more massive at CC in model 1 compared to model 3 \& 4. The larger network size in model 4 results in a more compact C-O core (yellow and green solid lines) compared to model 3. However, the final sizes of the Si and Fe cores remain somewhat less affected due to differences in network size.}


\subsection{Entropy from Oxygen Burning Onward}

\deleted{In Fig. \ref{fig:13M_26M_ye_evol}, we show the evolution of the electron mole fraction (number of electrons per nucleon) $Y_e= \Sigma_i X_i Z_i / A_i$ inside a mass coordinate of 2 \msun \ (approximately the boundary of the carbon core) of a 13 \msun \ and a 26 \msun \ star at various advanced nuclear burning stages. This is shown from the central oxygen ignition stage and subsequent depletion followed by silicon ignition and depletion until the core collapse (CC) stage. While initially the $Y_e$ was nearly uniform throughout the core, the central value of $Y_e$  steadily decreases until at CC, it reaches about 0.460 slightly more neutron rich than the nucleus \iso{56}{Fe}. Both 13 \msun \ and 26 \msun \ stars have very similar central $Y_e$ and $Y_e$ structures till the mass coordinate of $m_{interior}$ = 1.0 \msun \ at the core collapse stage, although the iron core mass at CC is slightly larger for the 26 \msun \ star (1.52 \msun \ instead of 1.48 \msun). Much of the reduction of $Y_e$ in the interior $\sim$ 1\msun \ takes place from Si burning onwards. The entropy at the inner edge of the Fe core at CC are also similar for both stars.
}


\deleted{As noted by \citet{Woosley:1972aa}, it is important to account for weak-interaction mediated reactions, in particular electron capture reactions early on 
e.g. during the oxygen burning stage onwards
 This is illustrated in Fig. \ref{fig:13M_26M_ye_evol} in the decrease of $Y_e$ between the time of oxygen ignition and depletion, especially inwards of the core mass of $\sim$ 0.94 $M_{\odot}$. The weak reaction network employed in (\texttt{mesa$\_$79.net}, which contains isotopes up to $^{60}Zn$) used here includes the rates calculated by \citet{Fuller:1982aa,Fuller:1982ab}, \citet{Oda:1994aa} and \citet{Langanke:2000aa}. The inclusion of accurate weak reaction rates together with finely resolved zoning of the core and appropriate evolutionary time stepping leads to small iron core masses with low entropies in the core prior to collapse \citep{Weaver:1985aa} - both are conducive towards stellar explosion rather than shock stalling and a fizzle. At the time of central Si exhaustion, the mass interior to the oxygen burning shell is 1.32 \msun . Subsequently, the core goes through a size contraction and increased temperature leads to the burning out of the oxygen shell to about 1.65 \msun \ at CC. The location of the oxygen burning shell ultimately determines the entropy profile which subsequently determines the extent of the silicon burning shell and thus the final iron core mass poised to undergo collapse under its own weight.
}

\begin{figure*}
\gridline{\fig{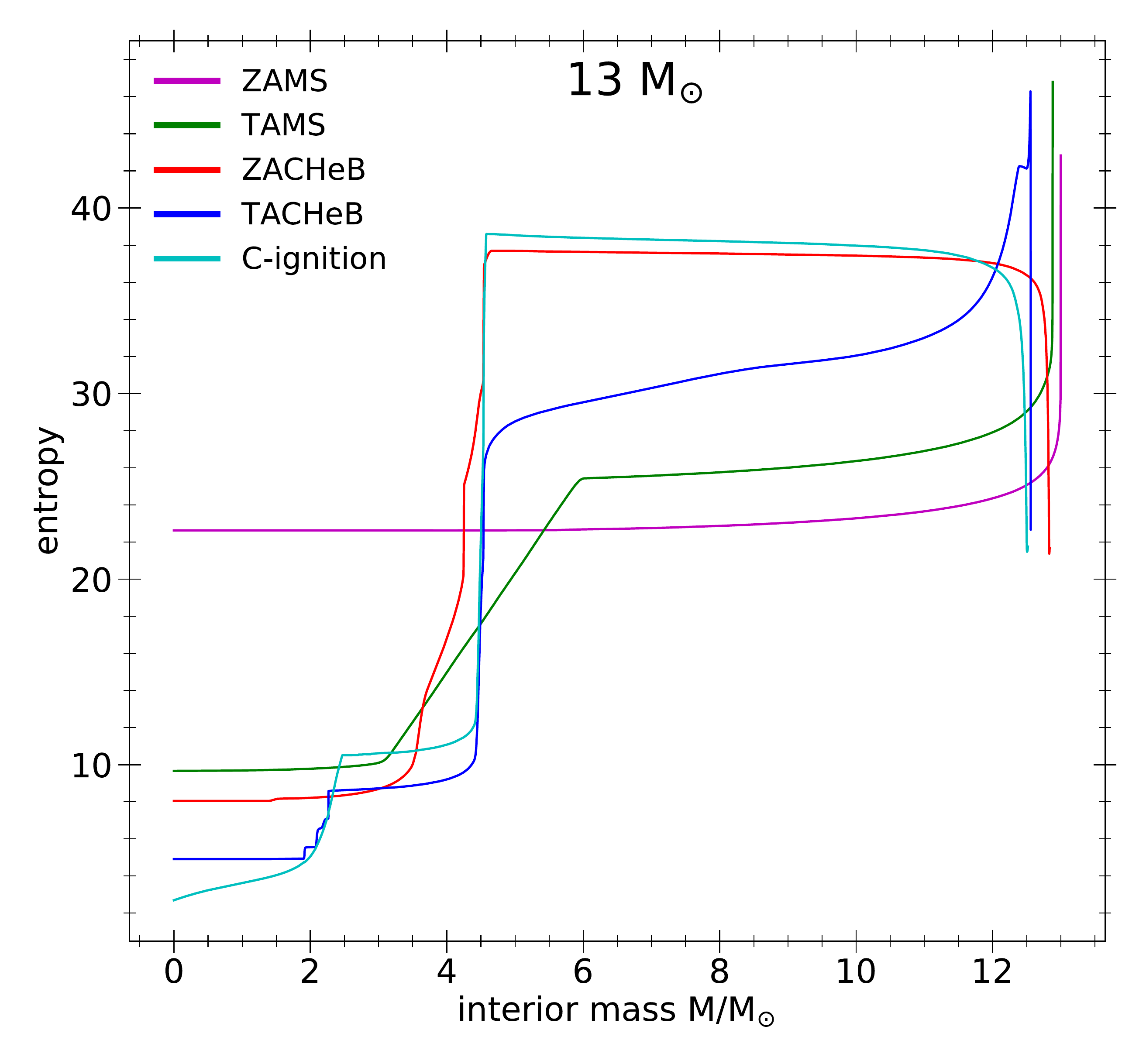}{0.45\textwidth}{(a)}
				\fig{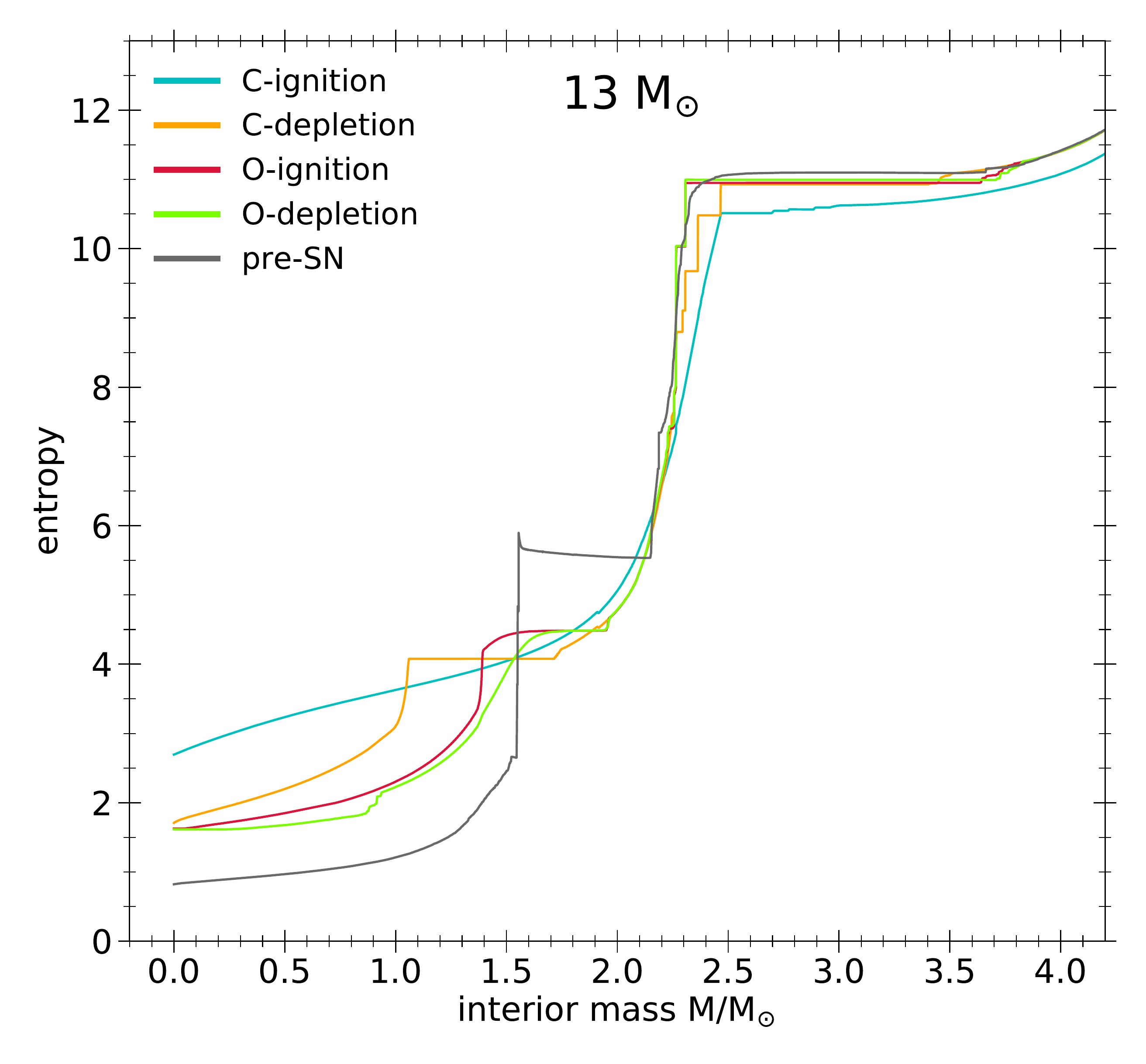}{0.45\textwidth}{(b)}}
\gridline{\fig{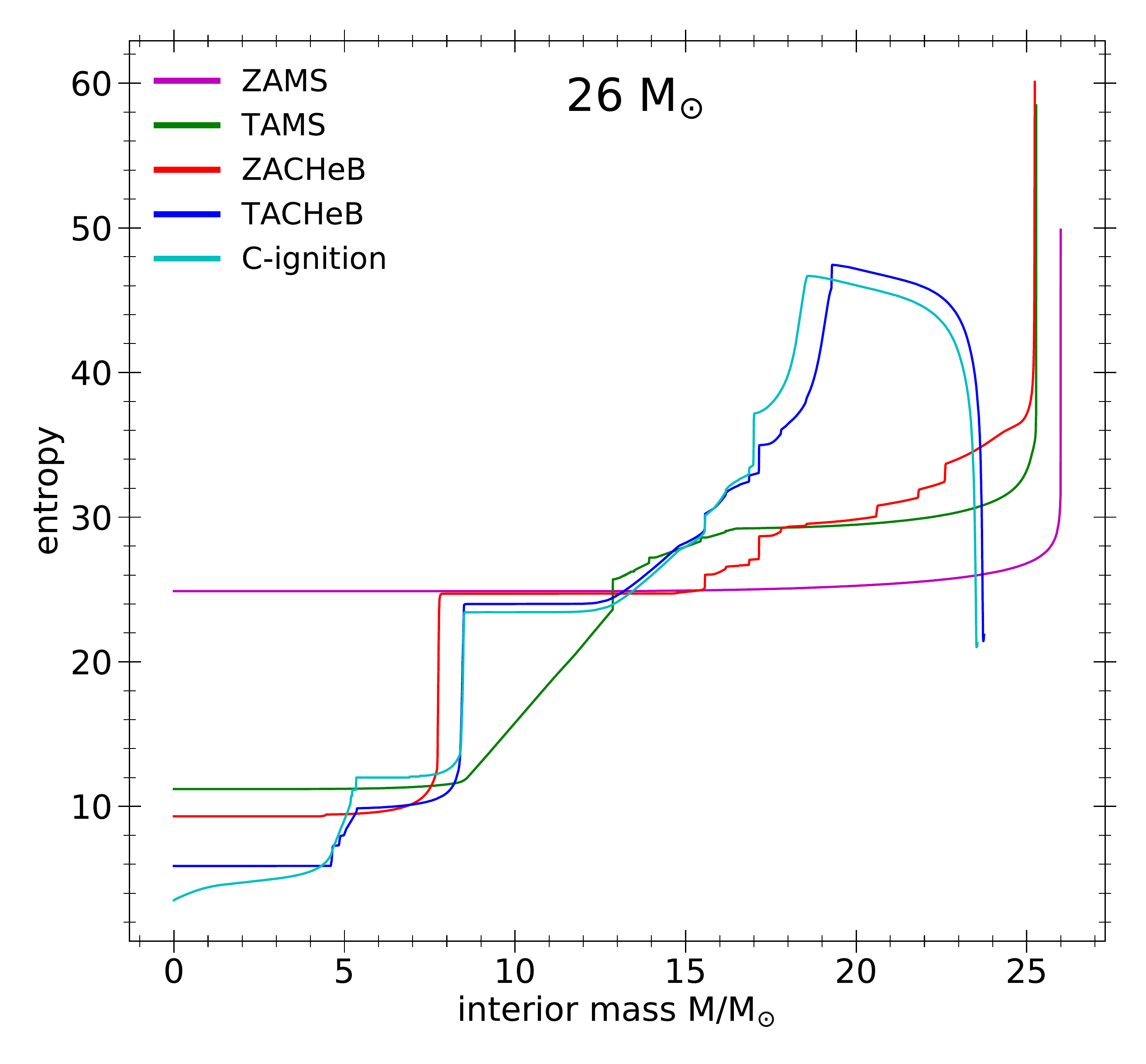}{0.45\textwidth}{(c)}
				\fig{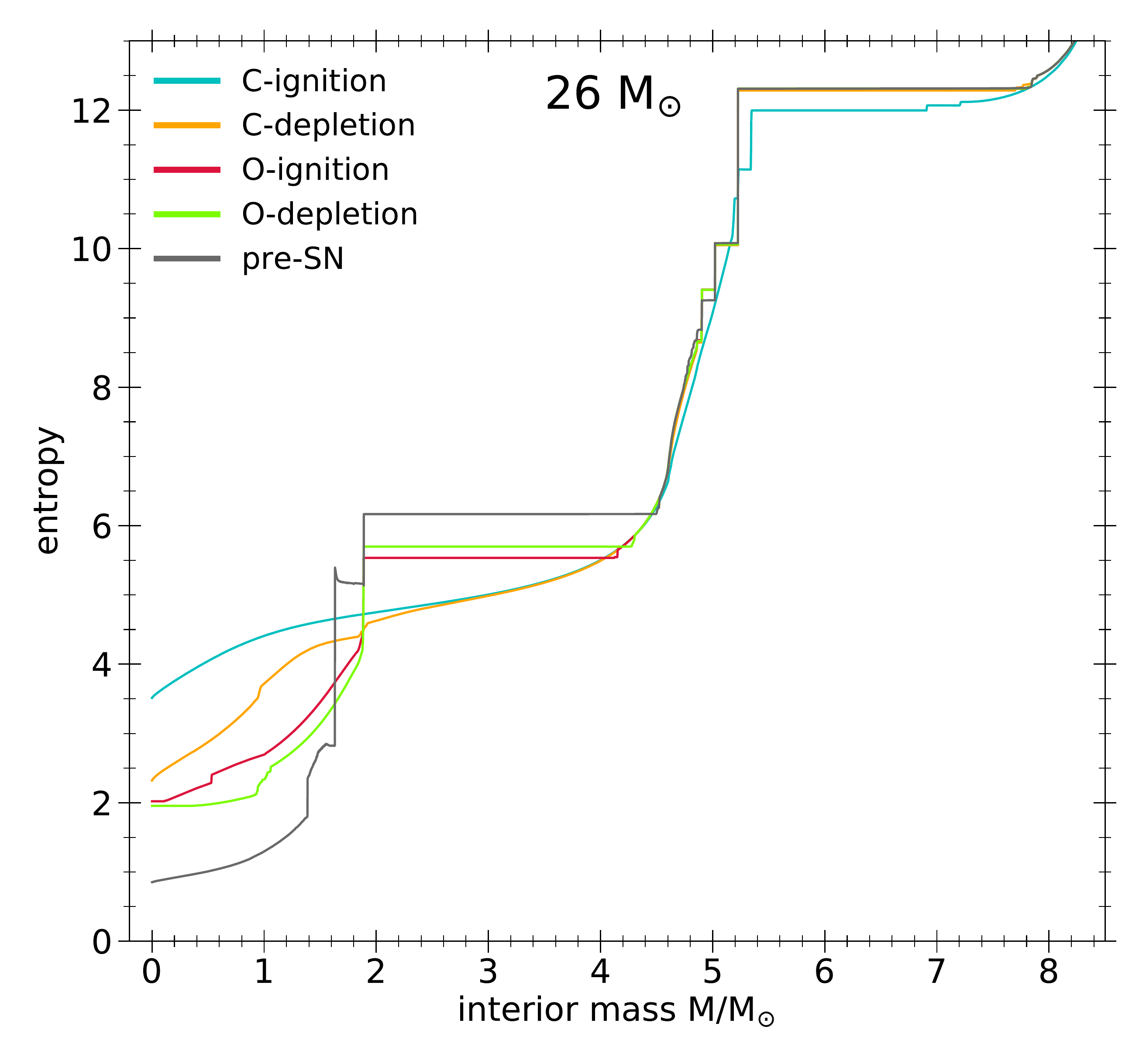}{0.45\textwidth}{(d)}}
\caption{Entropy profile evolution for the M$_{ZAMS}$= 13 \msun \ model 3 and 26 \msun model 1,  with Z = 0.006, described in Table \ref{tab:13m_26m_param}. The total entropy per baryon, S/N$_A$k is plotted as a function of mass co-ordinate in panel (a) for evolution from ZAMS through C-ignition and panel (b) from C-ignition through CC stages. The pre-SN entropy profile structure is a consequence of increased entropy due to losses due to neutrinos in late burning stages.
\label{fig:13M_26M_entropy_evol}}
\end{figure*}

The \deleted{resultant} entropy profiles at various stages are shown in Fig \ref{fig:13M_26M_entropy_evol}\deleted{ (right panel and}\added{, from ZAMS through C-ignition in panels (a) \& (c) and C-ignition through CC in panels (b) \& (d) for the two ZAMS masses. The profiles} show sharp increases at the boundaries of various fossil shell-burning zones. Due to high core temperatures, the core is cooled predominantly by neutrinos emitted by thermal processes rather than by radiation\deleted{ even} from carbon burning stage onwards. However, oxygen burning onwards, the non-thermal emission of neutrinos as a result of electron capture on nuclei and beta decay of others start contributing. \deleted{Since neutrinos once produced 
escape from the core without any further interaction, they are efficient cooling agents and cause the entropy to decrease in the central parts of the core. Sharp entropy gradients develop at the edges of the core (or in regions of shell burning) that extends upto where the convective transport keep the composition mixed and the temperature high.}\added{Neutrinos are efficient cooling agents as they escape from the core without any further interaction once produced. This causes entropy to decrease in the central parts of the core. At the edges of the core (or in regions of shell burning), sharp entropy gradients develop that extends upto the region which is well mixed by the convective transport and where the temperature is high.} These sharp entropy gradients that prevent the mixing and penetration of burning products lead to the onion-skin structure in the core with distinct elements (e.g. C, Ne, O, Si) \deleted{structures}. Panel (a) \deleted{of Fig \ref{fig:13M_26M_entropy_evol} shows the entropy evolution of a 13 \msun \ star from ZAMS till core carbon ignition stage. It } shows that as the star's entropy in the core decreases, the entropy in the envelope increases with time, even though during the various \added{early} nuclear burning stages the entropy profile remains flat for a large part of the star except near the surface of the star.\deleted{Panel (b)  on the other hand shows the interior entropy structure from the carbon ignition to the core collapse stage. Panels (b) \& (c) show similar plots for a 26 \msun \ star.} \added{Panel (c) shows similar trend for 26 \msun \ star.}

\subsection{Mass loss rate as the star evolves}

As discussed in the Methods section, we use a combination of mass loss rates calculated for hot stars as in \citet{Vink:2001aa} appropriate for OB stars near the main sequence and \deleted{for other cooler stars we use rates of \citet{de-Jager:1988aa}.}\added{a combination of rates calculated under the ``Dutch" scheme for cooler stars.} The rates for red supergiants have been tested recently by \citet{Mauron:2011aa} who found it consistent with the \citet{de-Jager:1988aa} rates. \citet{van-Loon:2005aa} however have considered RSGs believed to be dust-enshrouded and have argued for a much higher mass loss rate. \citet{Meynet:2015aa} pointed out that these mass loss rates at a given luminosity can differ by more than two orders of magnitude.


\deleted{In Fig. \ref{fig:13M_26M_mdot} we show the mass loss rate vs time before collapse of the star from ZAMS to core collapse for a variety of model runs summarized in Table \ref{tab:13m_26m_param} for both 13 \msun \ and 26 \msun \ stars. 
Although various internal parameters of the model evolution were tried out (namely different nuclear reaction networks, overshoot parameters and zoning resolutions) the mass loss rate (for the assumed $\eta_{Dutch}$ = 0.5 used in the runs) the mass loss rates appear very similar to each other, with somewhat small changes of the onset of transitions to high mass loss rate or from high to low mass loss rates.}

\begin{figure}[htb!]
\plotone{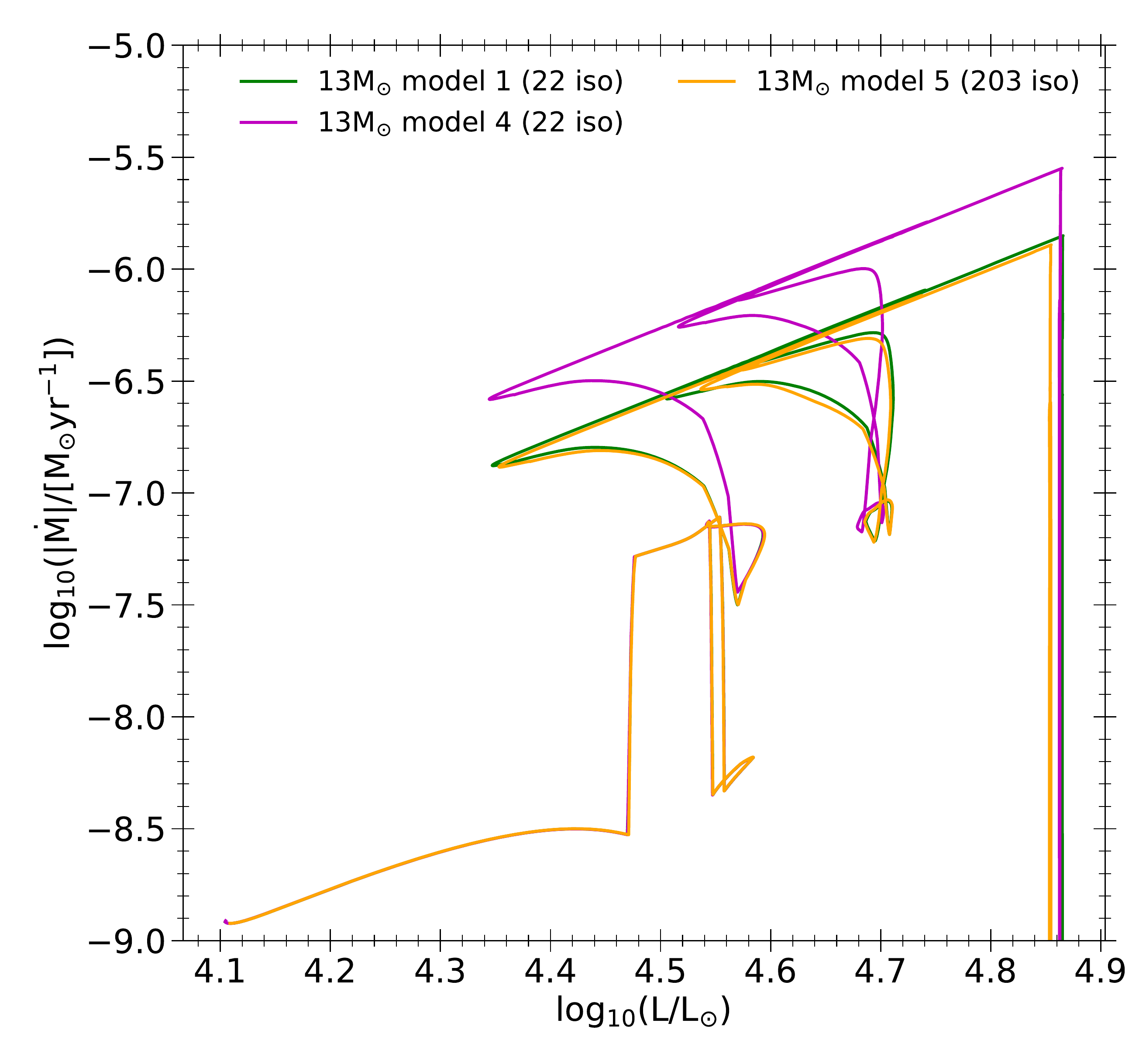}
\caption{The figure shows mass loss rate as a function of luminosity (left) and \Teff \ (right) from ZAMS for various M$_{ZAMS}$= 13 \msun \ , Z = 0.006  models described in Tables \ref{tab:13m_26m_param}. 
\label{fig:13M_mdot_L}}
\end{figure}

In Fig. \ref{fig:13M_mdot_L} we show the mass loss rate as a function of Luminosity 
for our fiducial 13 \msun \ \deleted{model 3}\added{models}.
\deleted{In Fig. \ref{fig:13M_26M_mdot} (a), the star in the ZAMS stage at the lower left corner,}\added{The star at ZAMS stage} has a low mass loss rate of \powten{-9} \msun \ yr$^{-1}$ which climbs to a factor of roughly three times higher value as hydrogen in its core is depleted and its color evolves towards the red. At a luminosity about \e{3 }{4} \Lsun \ (log L = 4.47) and a surface temperature of \Teff \ = 25,000 K, the mass loss rate undergoes a sharp increase by a factor of nearly 18, due to the so-called "bi-stability jump" \footnote{At the transition temperature the stellar mass loss rate, powered by the line-driving mechanisms, changes markedly due to recombination of dominant metal species and undergoes the mass loss rate jumps due to radiative acceleration in the subsonic wind part. In particular the bi-stability jumps are metal fraction (Z) dependent. Around \Teff \ = 25,000 K, Fe IV ions recombine to Fe III and as Fe III ions are comparatively more efficient line drivers, this leads to an increased line radiative acceleration and higher mass loss rate of the wind \citep{Vink:1999aa}.} \citep{Vink:2001aa}. The 26 \msun \ star \added{(plot not shown here)} also goes through the "bi-stability jumps" after the TAMS loop and about \powten{6} yr before CC, but because it stays as a RSG, its mass loss evolution is somewhat simpler compared to that of a 13 \msun \ star. 
The 13 \msun \ star crosses this \deleted{effective}\added{transition} temperature several times near the terminal age main sequence (TAMS) stage.\deleted{ and} As a result the mass loss rate undergoes both sudden upward as well as downward transitions within a short range of luminosities. 

In massive stars \deleted{with}\added{when} surface temperatures \deleted{falling}\added{fall} below 5000 K, dust begins to form in the stellar wind as the gas cools \deleted{with distance}\added{while it moves away} from the star. As the wind mass
loss is driven by stellar luminosity, and the mass loss rate is one of the factors that determines the amount of dust formed, the dust production rate was found to correlate with the
bolometric magnitude \citep{Massey:2005aa} as:
$$ log \; \dot M_{dust} = -0.43 M_{bol} - 12.0$$ 
\added{This relation is valid }for stars with $M_{bol} < -5$ that corresponds to stars more massive than $>10 \; M_{\odot}$. This \deleted{relation then}\added{therefore} gives a direct handle on how much dust formation is expected from a RSG of given luminosity at the end stage. The dust affects both stellar luminosity as well as color and its effects must be taken into account when comparing with observational color magnitude data. Dust formation and extinction due to dust are calculated in Appendix \ref{app:dust_ext} and the results for a 26 \msun \ star and the corresponding V and I band calculated magnitudes are reported in Table \ref{tab:dust_ext} \added{and discussed in the following section}.

\section{Summary and Discussion} \label{sec:conclusion}

\added{We have explored two ZAMS masses claimed in the literature for the progenitor star of SN 2013ej. We simulated the evolutionary stages up to the core-collapse of the progenitor star using MESA code. We used strict mass and temporal resolution controls recommended by \citet{Farmer:2016aa}. Our ZAMS 13 \msun \ and 26 \msun \ progenitor models have sufficient mass resolution and have successfully converged. We compared these models with the pre-SN observations by the HST about a decade before collapse \citep{Fraser:2014aa} and the visual extinction derived for the line of sight of the progenitor of SN 2013 ej and SN 2003gd \citep{Maund:2017aa}.}

\citet{Utrobin:2017aa} claim that a progenitor star with a ZAMS mass of 27.5 $\pm$ 2 \msun \ with an ejecta mass of 23.1 -- 26.1 \msun \ radius of 1500 \rsun \ at pre-SN stage explains better the initial peak of the light curve. In addition, they argue that the bipolar $^{56}$Ni distribution gives a better fit to the observed photospheric velocity profiles in their models. They prefer the 26 \msun \ ejecta model over the 23 \msun \ ejecta model. However, they do not give any explicit details about their pre-SN models, especially about the various physical parameters that would lead to such a progenitor star. We have run several MESA simulations to show that a star with ZAMS mass of 26 \msun \ will end up with a much smaller pre-SN mass of 22 -- 23.5 \msun \ (including the collapsing core mass), even with a moderate mass loss rate ($\eta_{\rm Dutch}$ = 0.5) with the reasonable choices for other parameters for the evolution of the star. Our 26 \msun \ progenitor has a pre-SN radius of $\sim$1000 \rsun ,\added{ substantially smaller than what is taken by \citet{Utrobin:2017aa},} while our 13 \msun progenitor has a pre-SN radius of $\sim$660 \rsun . We also find that a higher ZAMS mass star with the same mass loss scheme produces even smaller pre-SN progenitor \citep[see also fig. 3 of][which shows a similar trend]{Ugliano:2012aa}. In addition, a larger ZAMS mass (typically above 30 \msun) would enter the Wolf-Rayet regime, reducing its chances to explode as a type II supernova.

As seen in Fig \ref{fig:13M_26M_HR}, such a massive star would be highly luminous (L$_{star}$/\Lsun \ $\sim$ a few times \powten{5}) for most of its life before SN explosion. Such a star would require considerable amount of dust in the CSM to obscure the high luminosity. Even so, an observation in K band would not be affected by dust extinction. Unfortunately, we do not have any pre-explosion K band observations for the progenitor of SN 2013ej. As discussed in the Results section, our 26 \msun \ model 2 shows a large variation in luminosity \added{spirals in HRD} in the final year before collapse, which results in both V \& I band variation of \added{as large as} 0.3 mag. \added{These spirals if observed may indicate that the progenitor star is approaching the CC stage.} This variation \deleted{would be easily observable and} would not be affected by dust \deleted{either}\added{extinction}.  \deleted{But}\added{However,} such observations are not available for the progenitor of SN 2013 ej. Nevertheless, there exist HST observations for the progenitor star taken at about 8 \& 10 years before the explosion. We calculate the extinction values using the CSM profiles calculated from the output of our MESA simulation for our fiducial ZAMS 26 \msun \ model 1 as discussed in Appendix \ref{app:dust_ext}. These values are tabulated in Table \ref{tab:dust_ext}. To dim a star as bright as a 26 \msun \ star, the amount of extinction required A$_{\rm V}$ would be of the order of 3 mag.\deleted{ Our calculations for A$_{\rm V}$ range from 1.8 to 6 mag (if we consider dust consisting of both the Graphite \& Silicate grains). However, the higher values \added{of extinction} calculated using minimum distance for grain formation determined from adiabatic cooling of the gas is not physically appropriate. Considering the high radiation environment in the vicinity of the star, instead of using adiabatic cooling, a more appropriate way to evaluate the formation of dust and the resulting extinction it would lead to, would be to use the heat balance equation. In addition, the dust destruction temperature of 2000 K considered for the graphite is on the higher side. If we consider the dust destruction temperature of 1500 K for the graphite (column (7) in Table \ref{app:dust_ext}), then we end up with A$_{\rm V} \approx$ 2.} Even \added{though our calculations of the A$_{\rm V}$ values based on the CSM profiles allow such a high extinction, such }\deleted{this} value of A$_{\rm V}$ is \deleted{significantly higher than the value of}\added{excluded by the observed value of} 0.45 $\pm$ 0.04 determined by \citet{Maund:2017aa} in the host for SN 2013ej using archival pre-explosion observations through Bayesian analysis of stellar population. They also determined a value of A$_{\rm V}$ = 0.46$^{+0.04}_{-0.03}$ for another type IIP SN 2003gd, which occurred in the same host galaxy.  As this galaxy is observed face-on, most of this extinction would be due to the dust formed in the CSM around the progenitor star. This \deleted{diminishes the possibility of having}\added{poses a difficulty for} a ZAMS mass of 26 \msun \ for the progenitor star of SN 2013ej, \added{as advocated by \citet{Utrobin:2017aa}.}

\deleted{Although we consider}\added{The multi-wavelength observations of the SN, observationally derived pre-explosion dust extinction values, as well as our model results point towards a ZAMS mass of 13 \msun \ for} the progenitor of SN 2013ej\deleted{ to be better represented by a 13 \msun \ star} than a star of about twice its mass\deleted{, we are in the process of making}\added{. In our future paper, we will discuss} detailed simulations of post-explosion multi-waveband light curves and Fe line velocity evolution \added{of the two stars considered here} to compare with available observational data \added{for additional scrutiny of the mass of the progenitor star}.

\appendix

\section{Extinction by Circum-Stellar Material} \label{app:dust_ext}

Here, we discuss our calculations for the extinction by dust particles (and the underlying assumptions) in the CSM formed by the mass lost in stellar winds (values listed in Table \ref{tab:dust_ext}). The dust particles can form only when the wind moves sufficiently far from the star to avoid dust particles being eradicated by radiation near the star, and cools down to a temperature below dust destruction temperature. The typical value used in the literature for the sublimation temperature at which the dust particles will be destructed is $T_{d,max}$ = 1500 K \citep{Das:2017aa,Kochanek:2012aa}; however, the graphite dust particles can withstand temperatures as high as 2000 K \citep[see, e.g.,][]{Fox:2010aa}. We calculate the minimum distance ($R_{min}$) from the stellar surface at which the dust can form, where the temperature of CSM falls below $\rm T_{d,max}$. \citet{Das:2017aa} determined this distance by simple adiabatic cooling of the wind ($\rm TV^{\gamma-1} = constant$) due to expansion with an adiabatic monotonic gas ($\gamma = \frac{5}{3}$). In this case, the minimum distance the gas has to travel to cool down below the graphite destruction temperature, considering dust temperature is same as that of the surrounding gas, is given as
\begin{equation} \label{eqn:Rmin_adia}
R_{min}  = R_{star} \sqrt{\frac{T_{star}}{T_{d,max}}}
\end{equation}

We use a more robust way of calculating the minimum distance using heat balance equation for dust grains in thermal equilibrium, where the rate of dust heating due to stellar radiation is equal to the rate of dust cooling due to thermal emission, written as \citep[see, e.g.][]{Draine:1984aa,Kruegel:2003aa}
\begin{equation} \label{eqn:heat_bal}
\int_0^{\infty} Q_{abs}(a, \nu) \; J_{\nu,\; star} \; d\nu = \int_0^{\infty} Q_{abs}(a, \nu) \; B_{\nu}(T_d) \; d\nu
\end{equation}
The absorption efficiency $Q_{abs}$ in mid-infrared where the grain emission takes place can be approximated to a power-law using a simple spherical grain model \citep{Kruegel:2003aa}:
\begin{equation} \label{eqn:abs_eff}
Q_{abs}(a, \nu) = a \; Q_0 \; \nu^{\alpha}
\end{equation}
where, $Q_0$ is a constant. The value of $\alpha$, the emissivity index, ranges from 1 to 2; however, the value of 2 is favored in the literature. The R.H.S. of equation \ref{eqn:heat_bal}, which is the cooling rate, $\mathcal{W}$ ($\rm erg~cm^{-2}~s^{-1}$), then becomes 
\begin{eqnarray} \label{eqn:cooling_rate}
\mathcal{W} & = & \int_0^{\infty} Q_{abs}(a, \nu) B_{\nu}(T) d\nu \nonumber \\
& = & \frac{2h}{c^2} \ a Q_0 \int_0^{\infty} \frac{\nu^{3+\alpha}}{e^{h\nu /kT}-1} d\nu \nonumber \\
& = &\frac{2h}{c^2} \ a Q_0 \ \left(\frac{kT}{h}\right)^{4+\alpha} \int_0^{\infty} \frac{y^{3+\alpha}}{e^{y}-1} dy
\end{eqnarray}
The integral in above equation yields approximate values \citep{Kruegel:2003aa}:
\begin{eqnarray}
\int_0^{\infty} \frac{y^{3+\alpha}}{e^{y}-1} dy \simeq \left\{
\begin{array}{ll}
   								6.49, & \text{ if $\alpha$=0 } \\
    								24.89, & \text{ if $\alpha$=1 } \\
   							   122.08, & \text{ if $\alpha$=2 } \\
\end{array} \right.
\end{eqnarray}
Then with favored value of $\alpha$=2, equation \ref{eqn:cooling_rate} becomes \citep{Lequeux:2005aa}
\begin{equation} \label{eqn:RHS}
\mathcal{W} = 4.6 \times 10^{-11} \frac{a}{0.1 \mu m} T^6 \quad \text{(erg cm$^{-2}$ s$^{-1}$)}
\end{equation}
If we assume that the emitting dust is highly absorbing in UV and the visible, which are the only relevant wavelengths, then $Q_{abs}(a, \nu (UV)) \simeq$ 1. In this case, the L.H.S. of equation \ref{eqn:heat_bal} will be simply equal to $L_*/4\pi R^2$, where $R$ is the distance of the dust grain from the star. Combining this with equations \ref{eqn:heat_bal} and \ref{eqn:RHS}, we get the minimum distance at which dust grain of size $a$ with temperature $T_{d, max}$ could exist as 
\begin{equation} \label{eqn:Rmin}
R_{min}[cm] = \sqrt{\frac{L_* [erg~s^{-1}]}{4\pi \times 4.6 \times 10^{-11} \ T_{d,max}^6 \ (a/0.1 \ \mu m)}} 
\end{equation}
The ejected mass closer to star than this distance will not be able to form dust grains.

On the other hand, the mass that has moved too far away from the star will be too dilute to contribute to the extinction. We calculate the visual extinction assuming only the contribution by the dust particles existing in the CSM between $R_{min}$ and a maximum distance of $R_{max} \approx$ \powten{15} cm \cite[similar to][]{Utrobin:2017aa}, beyond which it will be part of the interstellar medium (ISM), and perhaps too dilute to contribute to the extinction value significantly for our models. The visual extinction can be calculated for a distribution of grains between a minimum and maximum grain size ($a_{min}$ \& $a_{max}$) using equation \citep{Perna:2003aa}
\begin{equation} \label{eqn:extinction_curve}
A_V = \int_{R_{min}}^{R_{max}} dr \sum_{i} \int_{a_{min}}^{a_{max}} da \; \pi a^{2} \: \frac{dn_i}{da} \: Q_{abs,i}(a,V)
\end{equation}
where, we neglect the angle dependent term involving $Q_{scat}(a,\nu)$ as MESA is a 1-D code. The size distribution $dn_i/da$ of the dust grain of type $i$, per unit volume per grain size $a$ per H-atom (in units of $cm^{-4}$) typically follows s a MRN \citep{Mathis:1977aa} power law given as
\begin{equation} \label{eqn:size_dist}
\frac{dn_i}{da} =A_{i} \; n_{H} \; a^{-\beta}, \; a_{min} \leq a \leq a_{max}
\end{equation}
where, the typical values for the ISM for $a_{min}$, $a_{max}$ and the index $\beta$ are 0.005 $\mu$m, 0.25$\mu$m and 3.5, respectively. This distribution would change for extreme conditions close to a highly luminous RSG star. Due to the radiation from a star, the larger grains are fragmented into smaller grains, while the much smaller grains are easily sublimated away. This leads to a flatter grain size distribution over a narrower range of grain sizes. Following \citet[][fig.3]{Perna:2003aa}, we see that after a considerable amount of time, there would exist only graphite grains distributed between grain sizes $a_{min}$ and $a_{max}$  of 0.15 $\mu$m and 0.22 $\mu$m, respectively, and with $\beta \approx$ 1.4. The coefficients $A_{i}$ in equation \ref{eqn:size_dist} are related to the dust-to-gas ratio (by mass), $f_d$ \citep[see][]{Laor:1993aa, Perna:2003aa}. For dust consisting of only graphite, we have
\begin{equation} \label{eqn:mass_frac}
f_{d} = \frac{m_{dust}}{m_{gas}} = \frac{4 \pi a_{max}^{4-\beta}}{3 m_H(4-\beta)}\left[1-\left(\frac{a_{min}}{a_{max}}\right)^{4-\beta}\right] A_{gra}\rho_{gra}
\end{equation}
%
where, $\rho_{gra} \approx 2.26 ~g ~cm^{-3}$ is the density of graphite grains \citep{Draine:1984aa} and $m_H$ is mass of a hydrogen atom. We conservatively assume that only about 50\% of the available $^{12}$C has formed dust. We determine $f_d \sim$ a few times \powten{-4}, using the surface composition of the star. We find that the values of $Q_{abs}(a,V)$ are somewhat independent of $a$ for the above range of the grain sizes. We used an averaged value for $Q_{abs}(V)$ = 1.5 in this range using data provided online by \citet{Laor:1993aa}. Within this formalism, we can now reduce equation \ref{eqn:extinction_curve} to the form
\begin{equation} \label{eqn:A_nu}
A_V = \frac{\pi A_{Gra} \; Q_{abs}(V)}{m_H} \left[ \frac{a^{(3-\beta)}}{(3-\beta)}\right]_{a_{min}}^{a_{max}} \int_{R_{min}}^{R_{max}} dr \ \rho(r)
\end{equation}
where $\rho(r)$ is the CSM density profile calculated by using the mass loss by stellar winds with a constant speed of 10 $km~s^{-1}$ \citep[see equation 4 of][]{Das:2017aa}. We calculated $A_V$ using $R_{min}$ determined from the adiabatic cooling of the gas as in equation \ref{eqn:Rmin_adia} or the heat balance equation for as in equation \ref{eqn:Rmin} up to $R_{max}$. While choosing $R_{min}$ for heat balance equation, which depends on the grain size, we chose the distance corresponding to $a_{min}$. The values of the extinction in I band, $A_I$, were calculated using relation given by \citet[A$_{\rm I}$/A$_{\rm V}$ = 0.479]{Cardelli:1989aa}. 

We also calculated the visual extinction values using equation (4) of \citet{Walmswell:2012aa}, where they assume that the CSM dust consists of low-density silicates ($\rho_{sil} = 2.5~g~cm^{-3}$) for comparison, using our CSM profiles. We used the surface composition of the star to find the mass of the dust. We considered the elements that constitute silicate dust grains. To find the $R_{min}$ value we used the heat balance formalism explained above, with $T_{d,max}$ = 1500 K, which is typical for silicates. We used the same value of $R_{max}$ (\powten{15} cm) from the surface of the star as before. These extinction values for V and I bands are listed in Table \ref{tab:dust_ext} and discussed in the Results (section \ref{sec:results}).

\section*{Acknowledgments}
\acknowledgments
We thank the directors and the staff of the Tata Institute of Fundamental Research (TIFR) and the Homi Bhabha Center for Science Education (HBCSE-TIFR) for access to their computational resources. This research was supported by a Raja Ramanna Fellowship of the Department of Atomic Energy (DAE), Govt. of India to Alak Ray and a DAE postdoctoral research associateship to GW. GW thanks Rob Farmer for valuable feedback through private communication. The authors thank the anonymous referee for his/her constructive comments. The authors thank Ajay Dev, Viraj Meruliya students of the National Initiative of Undergraduate Science (NIUS) program at HBCSE (TIFR) .
The authors acknowledge the use of NASA's Astrophysics Data System.

\software{MESA r-10398 \citep{Paxton:2011aa,Paxton:2013aa,Paxton:2015aa,Paxton:2018aa}, Anaconda Spyder (Python 3.6)}

\bibliography{my_biblio.bib}




\end{document}